\documentclass[fleqn,usenatbib]{mnras}

\usepackage[T1]{fontenc}

\DeclareRobustCommand{\VAN}[3]{#2}
\let\VANthebibliography\thebibliography
\def\thebibliography{\DeclareRobustCommand{\VAN}[3]{##3}\VANthebibliography}

\usepackage{graphicx}	
\usepackage{amsmath}	
\usepackage{amssymb}	
\usepackage{pdflscape}	
\usepackage{newtxtext,newtxmath}

\title[Be-HMXBs as progenitors of ULXs]{High-Mass X-Ray Binaries with Be Donors as Ultraluminous X-Ray Sources}

\author[S. Karino]{
Shigeyuki Karino,$^{1}$\thanks{E-mail: karino@ip.kyusan-u.ac.jp}
\\
$^{1}$Faculty of Science and Engineering, Kyushu Sangyo University, 2-3-1 Matsukadai, 
Fukuoka 813-8503, Japan
}

\date{Accepted XXX. Received YYY; in original form ZZZ}

\pubyear{2015}

\begin{document}
\label{firstpage}
\pagerange{\pageref{firstpage}--\pageref{lastpage}}
\maketitle

\begin{abstract}

Since the detection of X-ray pulses from ultraluminous X-ray sources (ULXs) in 2014, neutron stars have been considered as their central objects.
However, it remains unclear how neutron stars can be brighter than the Eddington luminosity, and no unified view exists on the magnetic field of neutron stars and the degree of beaming.
Recent observations suggest that some X-ray pulsating ULXs have Be-type donors, and some of them occupy the same region as Be-type high-mass X-ray binaries (Be-HMXBs) on the Corbet diagram, which reveals the relation between spin and orbital periods.
This suggests that at least some ULXs are special cases of Be-HMXBs.
In this study, we use the framework of mass-accretion models for Be-HMXBs to investigate the conditions under which neutron stars achieve mass-accretion rates beyond the Eddington limit and become observable as ULXs.
We show that a Be-HMXB may become a ULX if the magnetic field of the neutron star and the density of the Be disc meet certain conditions.
We also show that, although a stronger magnetic field increases the brightness of a neutron star ULX with a Be donor, its brightness cannot exceed the Eddington limit by a more than a factor of ${\approx} 50$.
Finally, we propose a scenario whereby some normal Be-HMXBs may evolve into ULXs as the donor evolves into a giant.

\end{abstract}




\section{Introduction}

Ultraluminous X-ray sources (ULXs) are X-ray sources located outside the galactic centre with luminosities above the Eddington luminosity of stellar-mass objects.
A current controversy involving ULX sources debates whether they are stellar-mass black holes (BHs) with mass-accretion above the Eddington limit or intermediate-mass BHs with sub-Eddington mass-accretion rates~\citep{kaaret2017}.
However, the detection of X-ray pulses from ULXs in M81 revealed that some ULX sources are in fact rotating neutron stars (NSs)~\citep{bachetti2014}.
Since their first detection, several ULXs have been found to emit X-ray pulses, and it is now widely recognised that NSs constitute a certain fraction of ULXs~\citep{carpano2018, doroshenko2018, furst2016, israel2017a, israel2017b, tsygankov2017, rodriguez2020}.
Despite the increasing number of NS ULXs being observed, it remains unclear how NSs emit X-rays far beyond the Eddington limit.
In particular, whether the magnetic field of ULX NSs is strong or weak remains hotly debated~\citep{eksi2015, tong2015, king2019, mushtukov2015, mushtukov2021, middleton2019, brightman2018, walton2018}.
In recent years, modelw that comprehensively treat spin, magnetic field and beaming of NS ULXs have been considered~\citep{erkut2020,king2020,mushtukov2021,abarca2021}.

\begin{figure}
\includegraphics[width=\columnwidth]{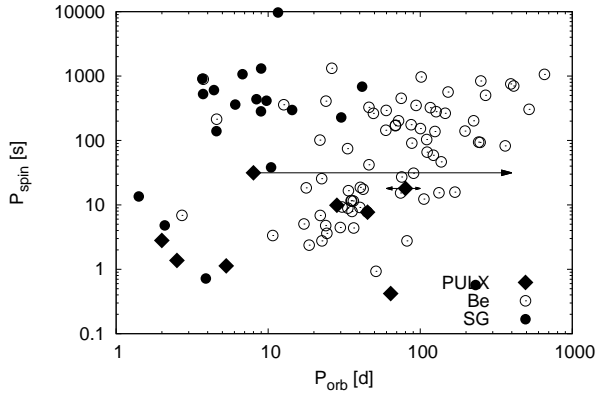}
\caption{
The Corbet diagram, which shows the relationship between orbital period and spin period of neutron stars.
HMXB systems are shown by open and solid circles, where the former show systems whose donors are supergiants and the latter show systems that have Be donors.
The black diamonds show the position of the ULXs harbouring neutron stars. The data were taken from~\protect\citep{coe2015, klus2014, martinez2017,king2020}.
}
\label{fig:CorbetDiagram}
\end{figure}

In addition, as ever more NS ULXs are observed, a major question has become the fraction of ULXs that are of NS origin versus BH origin \citep{fragos2015,shao2015,misra2020}.
To estimate the population of NS ULXs, we must understand the properties of binary systems in NS ULXs.
Based on current observations, NS ULXs appear to be a type of high-mass X-ray binary (HMXB) objects with a relatively massive companion (Table~\ref{table1} summarises the observed properties).
Some NS ULXs have supergiants as companions, whereas other systems appear to Be-HMXBs with Be-type companions (see Table~1 in~\citet{king2019}).
To study of HMXBs, Corbet diagrams are often used to show how the NS orbital period is related to the rotation period \citep{corbet1984, corbet1986}.
In such diagrams, some NS ULXs occupy the same region as Be-HMXBs (see Fig.~\ref{fig:CorbetDiagram}).
If the donor star is a supergiant and the system has a short orbital period, mass-transfer from the donor to the NS will occur via Roche lobe overflow (RLOF), which leads to a large mass-transfer rate~\citep{bildsten1997,frank2002}.
In this case, one can easily imagine that the mass-transfer rate exceeds the Eddington limit, allowing it to emit X-rays above the Eddington luminosity.
However, it is another matter to determine whether the material flowing into the NS can be accreted onto the NS without being lost by outflow~\citep{chashkina2019, kosec2018,lipunova1999}.
Conversely, the X-ray luminosity of Be-HMXBs is generally less than that of HMXBs with supergiant donors~\citep{bildsten1997}.
Be-HMXBs produce periodic X-ray outbursts with a period related to the orbital period of the Be-HMXB and sometimes produce giant outbursts that are abnormally bright~\citep{rivinius2013} (the X-ray luminosity of a typical normal burst is about $10^{36}\ \rm{erg \, s}^{-1}$, whereas a giant outburst it is over tenfold brighter).
An interesting question is therefore whether such binary systems can produce X-ray luminosities above the Eddington luminosity~\citep{okazaki2013, karino2016}; in other words, are some NS ULXs special cases of Be-HMXBs?

Given that Be-HMXBs represent a significant fraction of HMXBs~\citep{bildsten1997,coe2015}, a significant number of NS ULXs may exist since they follow part of the regular evolution of Be-HMXBs (even if no pulses are detected)~\citep{inoue2020, king2020, kuranov2020, mushtukov2021}.
The NS/BH fraction of the ULX population is important not only for calculating the initial mass function of massive stars but also for estimating the population of gravitational wave sources such as NS-NS binary stars~\citep{mondal2020}.

With these motivations, we examine in this study whether Be-HMXBs can indeed be ULXs.
For an NS binary system with a Be donor to be a ULX, it must satisfy the following criteria:
\begin{enumerate}
\item[(1)] The mass-transfer rate from the Be donor to the NS must exceed the Eddington limit.
\item[(2)] The material supplied from the donor must accrete onto the NS at a rate exceeding the Eddington limit.
\end{enumerate}
In this study, we divide the mass-accretion process from the Be donor to the NS according to criteria (1) and (2) above and investigate under what conditions Be-HMXBs can become ULXs.

In the next section, we discuss under which conditions a Be-HMXB can become a ULX.
Based on the conditions obtained in Section 2, we consider in Section 3 what type of binary system can become a ULX, discuss its evolutionary path, and compare the results with observations.
Finally, Section 4 is devoted to the conclusion.

\begin{table*}
	\centering
	\caption{\label{table1}
Parameters of NS ULXs. The Table uses the data from the Table of~\citet{king2019} and arranges the data as follows: (1)~\citet{israel2017a}, (2)~\citet{bachetti2014}, (3)~\citet{heida2019}, (4)~\citet{fragos2015},
(5)~\citet{rodriguez2020}, (6)~\citet{israel2017b}, (7)~\citet{furst2016}, (8)~\citet{wilson2018},
(9)~\citet{weng2017}, (10)~\citet{tsygankov2017}, (11)~\citet{Sathaprakash2019}, (12)~\citet{doroshenko2018}. ‘SG’ means ‘supergiant’ and ‘HG’ means `Hertzsprung gap.’	
	}
	\label{tab:Table}
	\begin{tabular}{lcccccl} 
		\hline
		Object name & $L_{\rm{X, max}}\ \rm{[erg \, s^{-1}]}$ & $P_{\rm{s}}\ \rm{[s]}$ & $P_{\rm{orb}}\ \rm{[d]}$ & Donor & Eccentricity
		& Reference \\
		\hline
		NGC5907 ULX-1 & ${\approx} 10^{41}$ & 1.13 & 5.3 & HG or SG? & & (1) \\
		M82 X-2 & $2 \times 10^{40}$ & 1.37 & 2.51 & ${>}5.2 M_{\odot}$ & & (2), (3), (4) \\
		M51 ULX-7 & ${\approx} 10^{40}$ & 2.8 & 2.0 & ${>}8.3 M_{\odot}$ & & (5) \\
		NGC7793 P-13 & $5 \times 10^{39}$ & 0.42 & 63.9 & $18 M_{\odot}\text{--}23 M_{\odot}$ SG & ${<} 0.15$ & (3), (6), (7) \\
		NGC300 ULX-1 & $4.7 \times 10^{39}$ & 31.5 & ${>}8$ & $8-10 / 40 M_{\odot}$ Be/SG & & (3), (8) \\
		SMC X-3 & $2.5 \times 10^{39}$ & 7.8 & 45 & ${>}3.7 M_{\odot}$ Be & 0.23--0.26 & (9), (10) \\
		M51ULX-8 & $2 \times 10^{39}$ & & & & \\
		NGC1313 PULX & $1.6 \times 10^{39}$ & 765.6 & & ${<} 12 M_{\odot}$ Be & & (11) \\
		Swift J0243.6+6124 & ${>}1.5 \times 10^{39}$ & 9.86 & 28.3 & Be/Oe & $ 0.1$ & (8), (12)\\
		NGC2403 ULX & $1.2 \times 10^{39}$ & 18 & 60--100 & Be & &\\
		\hline
	\end{tabular}
\end{table*}

\section{Binary Models and Mass-Transfer Processes in the Systems}

This section starts by giving the assumptions used for describing Be-NS binary systems, and then discusses the models.

\subsection{Binary Model}

\begin{figure}
\includegraphics[width=\columnwidth]{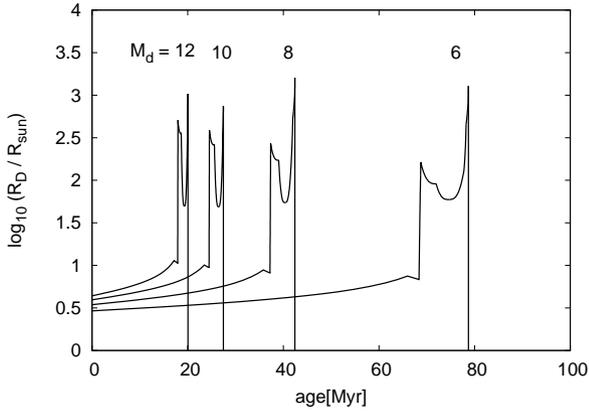}
\caption{
Donor radius. Time evolution of radii of Be stars in typical mass range ($6 M_{\odot}\text{--} 12 M_{\odot}$). Stars in this mass range have a radius of 4--10 solar radii during the main-sequence, although the radius increases rapidly above $10 R_{\odot}$ at the end of the main-sequence.
}
\label{fig:donorradius}
\end{figure}

In this study, the NS mass $M_{\rm{NS}}$ and radius $R_{\rm{NS}}$ are fixed at $1.4 M_{\odot}$ and 10 km, respectively.
In addition, we assume a Be-type donor star with mass $M_{\rm{d}} = 10 M_{\odot}$, which is almost the median value of the typical mass range of Be-type stars ($3M_{\odot} \text{--} 18 M_{\odot}$)~\citep{townsend2004}.
We further assume that the Be star ejects its mass through a deccretion disc.
The mass loss rate of a single $10 M_{\odot}$ B-type star in its main-sequence phase is estimated to be ${<} 10^{-8} M_{\odot}\ \rm{yr}^{-1}$~\citep{vink2001}, and even a Be star in a binary system is unlikely to exceed this rate by many orders of magnitude.
Therefore, the donor mass decreases only negligibly on the timescale of 10 million years, which is the typical main-sequence lifetime of a $10 M_{\odot}$ star.
Conversely, the donor radius $R_{\rm{d}}$ varies rather strongly during the main-sequence phase.
Figure~\ref{fig:donorradius} shows the approximate variation of the star radius over time, calculated as per~\citet{hurley2000}.
In this figure, the end of the main-sequence corresponds to the point where the stellar radius expands rapidly over a short time.
These results show that the stellar radius generally increases from ${\approx} 3 R_{\odot}$ to $10 R_{\odot}$ during the main-sequence phase.

The observed orbital period $P_{\rm{orb}}$ of Be-HMXB ranges from a few days to several hundred days, and the orbital eccentricity $e$ ranges from near-circular to highly distorted~\citep{townsend2011, klus2014, coe2015}.
In this study, we consider the orbital period and orbital eccentricity as the main parameters of a binary system and examine various combinations thereof over a wide range of parameter space.

The NS magnetic field $B$ is typically about $10^{12}\ \text{G}$ in most HMXB systems, as determined from their spin rates and cyclotron resonant scattering features~\citep{christodoulou2017}.
In contrast, the NS magnetic field of ULXs remains debated~\citep{eksi2015, tong2015, mushtukov2021}.
The NS magnetic field initially decays on the timescales of Ohmic decay, Hall dissipation, and mass-accretion~\citep{auguilera2008, zhang2006}.
Unfortunately, the age of NSs in NS ULXs, the initial magnetic field strength, and the cumulative accreted mass all remain unknown.
Recent studies argue that the magnetic field strength of NSs in ULXs are almost on a par with that of normal HMXBs: $B_{\rm{NS}} = 10^{11} \text{--} 10^{12}\ \rm{G}$~\citep{brightman2018, middleton2019, walton2018}.
Thus, we focus herein on these typical values and vary the magnetic field in several ways.

\subsection{Accretion Model}
\subsubsection{Mass-Transfer Rate}

In Be-HMXBs, the compact object receives mass from the Be stellar disc.
\citet{okazaki2013} shows that, in the case of a large outburst (type II outburst), the disc material is accreted in a Bondi-Hoyle-Littleton (BHL) process.
According to the BHL process, when an NS interacts with the disc material, it captures a part of the Be disc inside the accretion radius:
\begin{equation}
R_{\rm{acc}} = \frac{2 G M_{\rm{NS}}}{v_{\rm{rel}}^{2}}.
\label{eq:racc}
\end{equation}
In this case, the mass-transfer rate from the Be disc to the NS is~\citep{bondi1944, hoyle1939, edgar2004}
\begin{equation}
\dot{M}_{0} = \rho v_{\rm{rel}} \pi R_{\rm{acc}}^{2},
\label{eq:Mdot0}
\end{equation}
where $v_{\rm{rel}}$ is the relative velocity between the orbital velocity of the NS and the rotational velocity of the Be disc material, which varies depending on the inclination of the Be disc axis and the NS orbital axis.
$\rho$ is the density of the Be disc and is modelled as
\begin{equation}
\rho = \rho_{0} \left(\frac{D}{R_{\rm{d}}} \right)^{-7/2},
\label{eq:rho}
\end{equation}
where $\rho_{0}$ is the density of the inner edge of the Be disc.
The density of the Be disc and the rotational velocity of the disc material both vary with distance from the Be star.
However, we do not take into account the gradient of the physical quantities of the Be disc but represent them all at the periastron point $D = a(1-e)$, where $a$ is the semi-major axis of the orbit and can be obtained from Kepler’s law by specifying $M_{\rm{d}}$, $M_{\rm{NS}}$, and $P_{\rm{orb}}$.
In order to consider BHL accretion from a medium with a gradient of physical quantities, the integration of physical quantities within the accretion radius must be properly considered~\citep{wang1981,karino2019}.

The accretion of interstellar matter to a single NS is given by the mass-accretion rate (\ref{eq:Mdot0}); however, in a binary system, the gravity of the donor must also be taken into account.
As a result, the NS gravity exceeds donor gravity only inside the Roche radius:
\begin{equation}
R_{\rm{RL}} = D \frac{0.49 q^{2/3}}{0.69 q^{2/3} + \ln (1 + q^{1/3}) },
\label{eq:Rrl}
\end{equation}
where $q$ is the mass ratio~\citep{eggleton1983}.
Therefore, the mass-transfer rate from the Be disc to the NS neighbourhood in Be-HMXBs becomes
\begin{equation}
\dot{M}_{\rm{T}} = \min \left(1, \frac{R^2_{\rm{RL}}}{R^2_{\rm{acc}}} \right) \times \dot{M}_{0}.
\label{eq:MdotT}
\end{equation}
\citep{okazaki2013}.
However, if the Roche radius of the Be star is less than the donor radius at the periastron, instead of accretion from the disc, RLOF occurs from the donor itself.
In this case, a massive mass-transfer may occur, in which case the orbit would rapidly approach a circular orbit and/or, in some cases, a common envelope may form~\citep{eggleton2006}.
In such a case, the binary system would be unstable and may change rapidly, so the system would not be a steady ULX.

\subsubsection{Model of Accretion Disc around Neutron Star}

The material transferred from the Be disc to the NS forms an accretion disc and accretes onto the NS (such accretion discs have been studied in detail by~\citet{okazaki2013}).
If the accretion disc is a standard one~\citep{shakura1973}, the accretion proceeds on a viscous timescale, which is exceedingly long, so the material stored in the disc falls slowly into the NS.
In this case, even if a large amount of mass is transferred from the Be disc, accretion onto the NS is gradual, and the resulting X-ray curve will not be as bright and sharp as the observed giant outbursts seen from Be-HMXBs.
Therefore, the disc is likely to be a slim disc, which permits thermally stable, advection-dominated, supercritical accretion flow~\citep{abramovicz1988}.
To achieve supercritical accretion above the Eddington limit, the magnetic radius $R_{\rm{mag}}$ of the NS must be less than the photon-trapping radius $r_{\rm{trap}}$~\citep{okazaki2013}, which means that the mass-transfer rate exceeds a certain value.

Assuming the conditions given by~\citet{okazaki2013}, mass-accretion above the Eddington limit can be achieved only when the mass-transfer rate is supercritical:
\begin{equation}
\dot{M}_{\rm{T}} > \dot{M}_{\rm{sc}} = 1.7 \times 10^{19} k^{7/9} \mu_{30}^{4/9}.
\label{eq:Mdotsc}
\end{equation}
Here we use the normalised magnetic moment of the NS,
\begin{equation}
\mu_{30} = \frac{B R_{\rm{NS}}^{3}}{10^{30}\ \rm{G \, \rm{cm}}^3},
\label{eq:mu}
\end{equation}
where $k$ is a parameter related to the accretion geometry ($k=1$ for spherically symmetric accretion and $k=0.5$ for disc accretion).
Although the geometry of the slim disc may be intermediate between spherical and disc accretion, we use $k=0.5$ in this work.
Any system with a binary parameter set that does not satisfy the above condition for the mass-transfer rate [eq.~(\ref{eq:Mdotsc})] is rejected as a candidate for a NS ULX.

\subsubsection{Mass-Accretion Rate}
Material transferred from the Be disc to the neighbourhood of the NS will shift further through the slim disc to the vicinity of the NS if the condition (\ref{eq:Mdotsc}) is satisfied.
Eventually, when disc material is trapped in the magnetic field of the NS, it falls through the magnetic field lines and forms an accretion column near the magnetic poles, where it accretes and emits extremely bright X-rays~\citep{pringle1972}.
However, if the mass-transfer rate exceeds the Eddington limit, radiation pressure becomes dominant somewhere within the accretion disc, causing a significant thickening of the disc and outflows due to radiation~\citep{shakura1973, lipunov1982, lipunova1999, poutanen2007}.
The radius $R_{\rm{sph}}$ at which radiation pressure becomes dominant can be written as
\begin{equation}
R_{\rm{sph}} = 1.4 \times 10^{6} \frac{\dot{M}_{T}}{\dot{M}_{\rm{Edd}}}\ \rm{cm},
\label{eq:rsph}
\end{equation}
\citep{king2017}, 
where $\dot{M}_{\rm{Edd}}$ is the Eddington critical mass-accretion rate.
As matter accretes beyond this radius, its accretion rate decreases as
\begin{equation}
\dot{M}_{\rm{acc}} = \frac{R_{\rm{mag}}}{R_{\rm{sph}}} \dot{M}_{T},
\label{eq:MdotM}
\end{equation}
once it becomes trapped in the NS magnetic field.
Here, the magnetic radius $R_{\rm{mag}}$ is
\begin{equation}
R_{\rm{mag}} = \left[ \frac{\mu_{30}^{4} }{ 8 k^{2} G M_{\rm{NS}} \dot{M}_{\rm{acc}}^{2} } \right]^{1/7}.
\label{eq:rmag}
\end{equation}
%
%
%
%
%
%
%
Eqs.~(\ref{eq:MdotM}) and (\ref{eq:rmag}) give $\dot{M}_{\rm{acc}}$, which is the mass-accretion rate of NS ULX candidates 
with Be-type donors.
From this we obtain the luminosity
\begin{equation}
L_{\rm{X}} = \dot{M}_{\rm{acc}} c^2,
\label{eq:LX}
\end{equation}
which we can compare with the observed ULX luminosities.

Since $R_{\rm{sph}} \propto \dot{M}_{T}$, the mass-accretion rate onto the NS, given in eq.~(\ref{eq:MdotM}), does not depend on 
$\dot{M}_{T}$
and only depends on $R_{\rm{mag}}$.
After some calculations, the mass-accretion rate can be written as
\begin{equation}
\dot{M}_{\rm{acc}} 
\propto B^{4/9},
\label{eq:MB}
\end{equation}
which implies that the luminosity of the NS ULX is determined only by the magnetic field strength of the NS~\citep{lipunov1982}.

\section{NS ULX Model}

The previous section considered the mass-transfer process from the Be disc to the vicinity of the NS and calculated how much of the transferred material is accreted onto the NS.
The result indicates that the mass-transfer rate must exceed a certain level for super-Eddington mass-accretion to proceed in the accretion disc. 
In addition, the RLOF must be absent from Be stars to produce a stable ULX.
To summarise, to become a ULX, a NS that accretes mass from a Be-type donor must satisfy the following three conditions: 
\begin{eqnarray}
\dot{M}_{\rm{T}} &>& \dot{M}_{\rm{sc}},
\label{eq:cond1}
\\
\dot{M}_{\rm{acc}} &>& \dot{M}_{\rm{Edd}},
\label{eq:cond2}
\\
D &>& R_{\rm{RL}}.
\label{eq:cond3}
\end{eqnarray}
The luminosity of a NS when these conditions are satisfied is expressed as 
\begin{equation}
L_{X} = \frac{\dot{M}_{\rm{acc}}}{\dot{M}_{\rm{Edd}}} L_{\rm{Edd}}.
\label{eq:LXEdd}
\end{equation}

\subsection{Possible Neutron Star ULX Systems}

\begin{figure*}
\includegraphics[width=5.5cm]{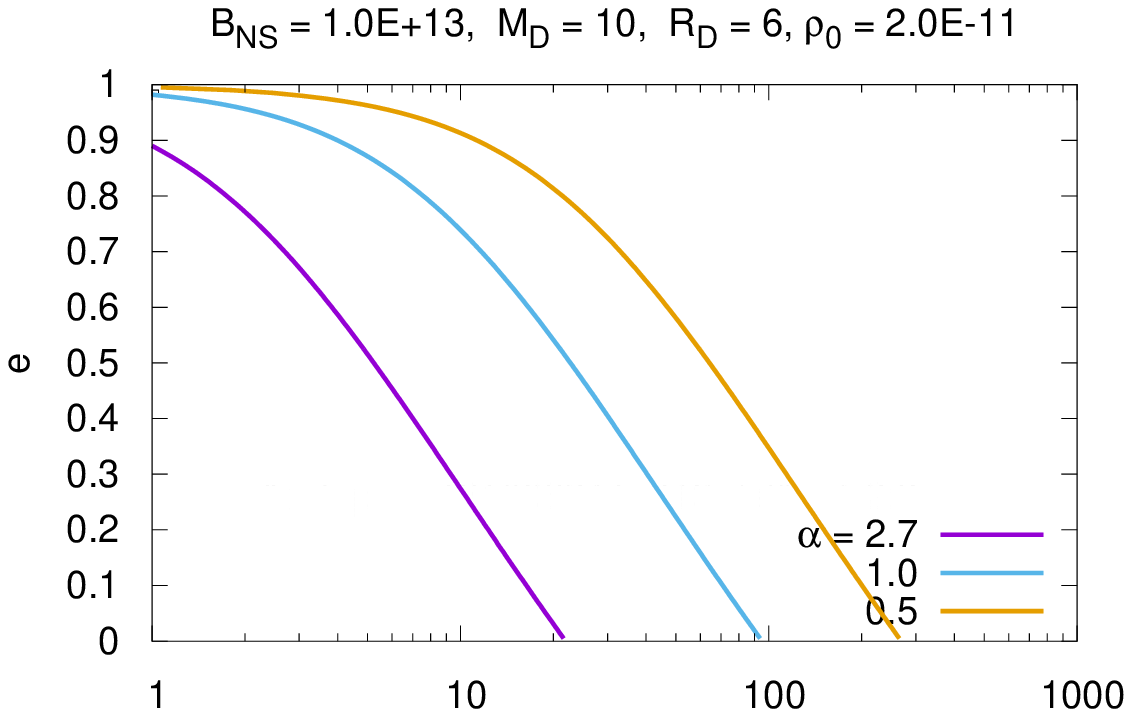}
\includegraphics[width=5.5cm]{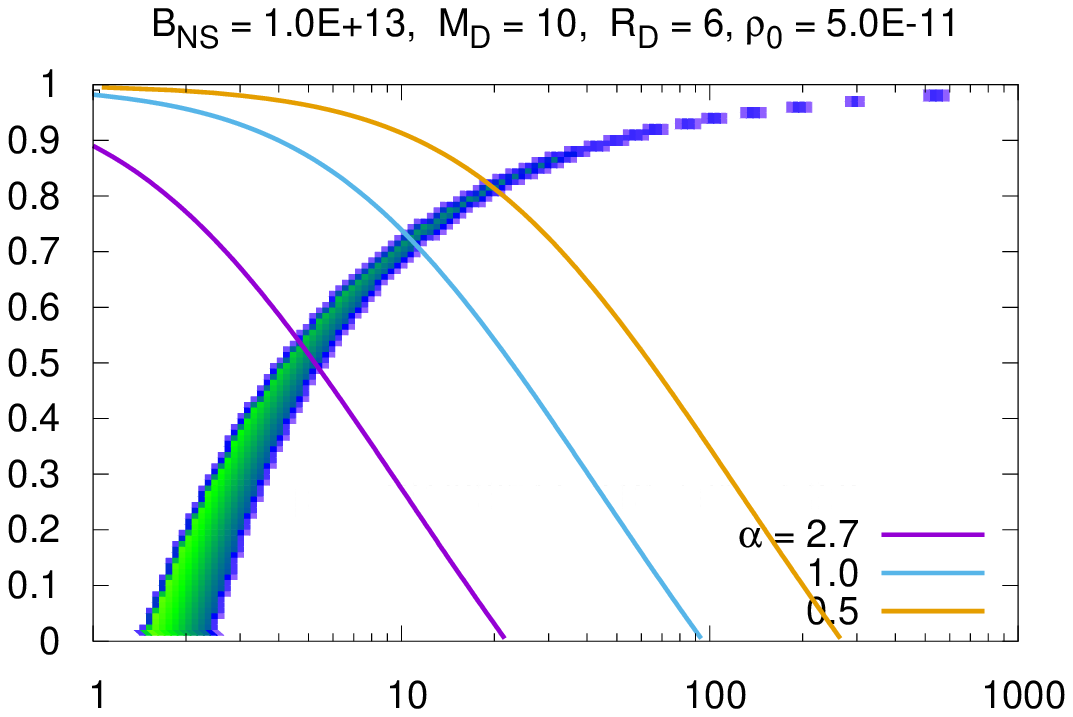}
\includegraphics[width=5.5cm]{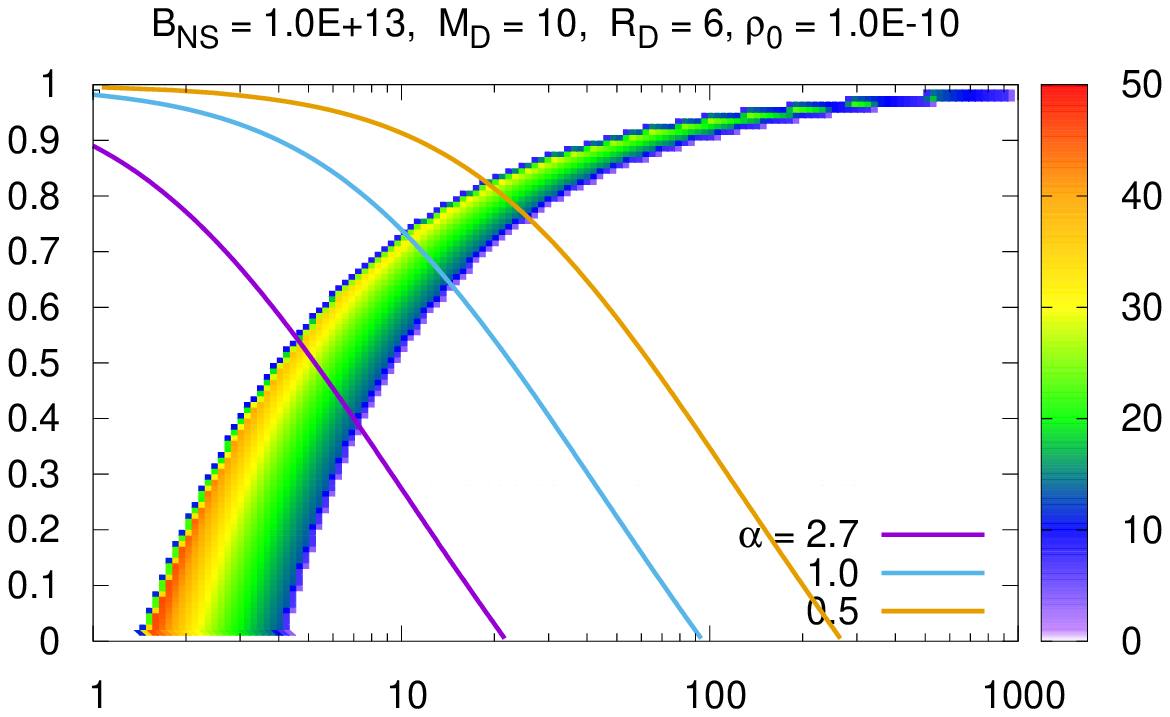} \\
\includegraphics[width=5.5cm]{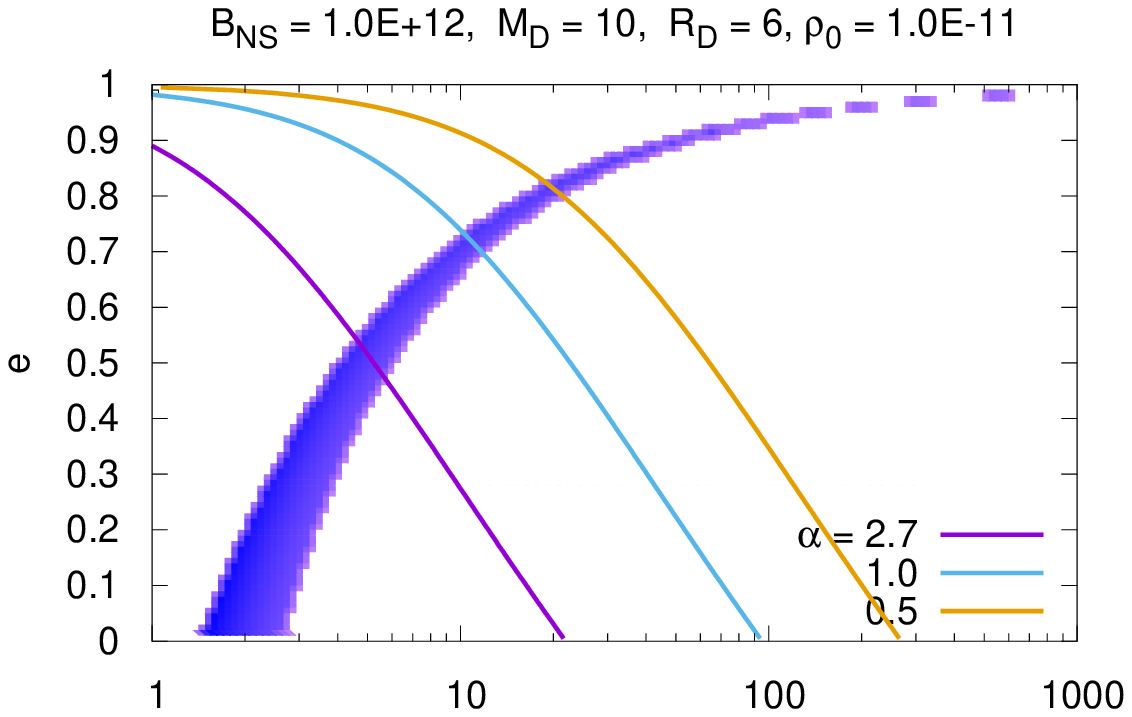}
\includegraphics[width=5.5cm]{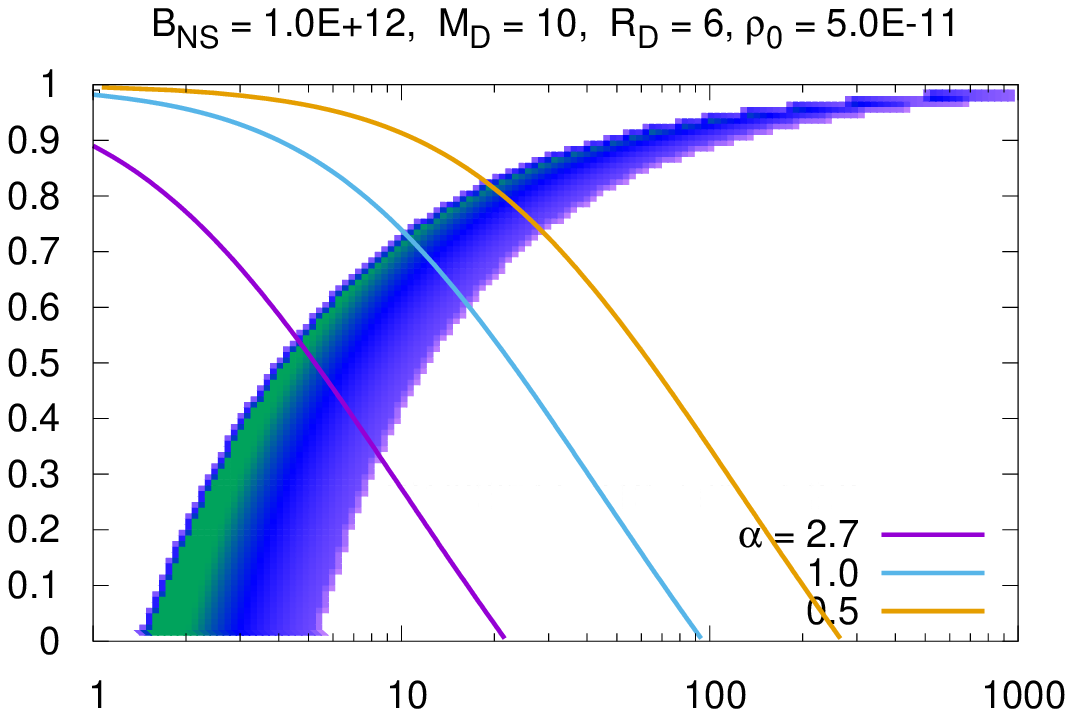}
\includegraphics[width=5.5cm]{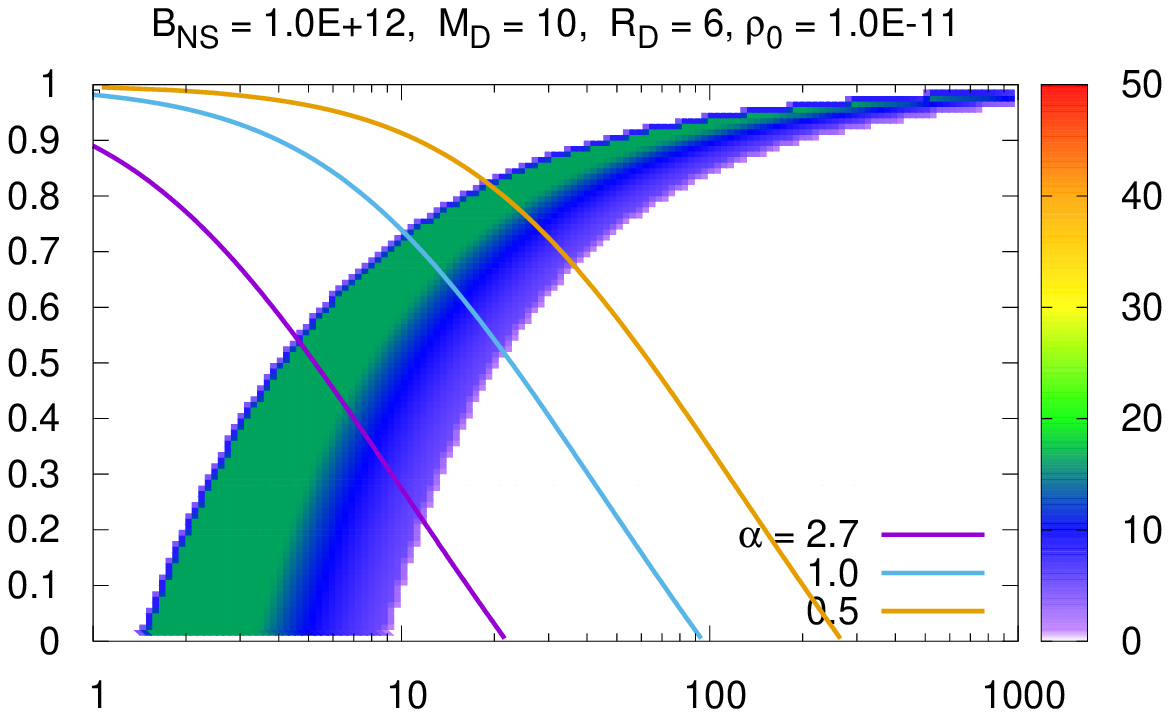} \\
\includegraphics[width=5.5cm]{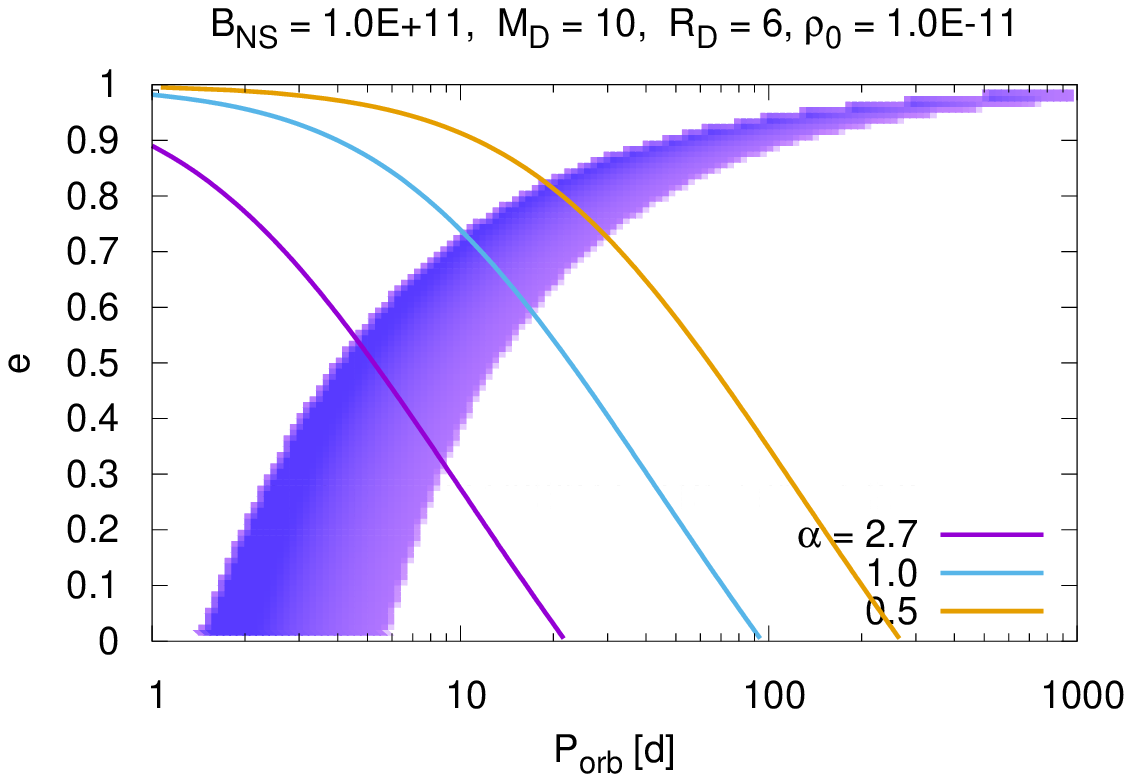}
\includegraphics[width=5.5cm]{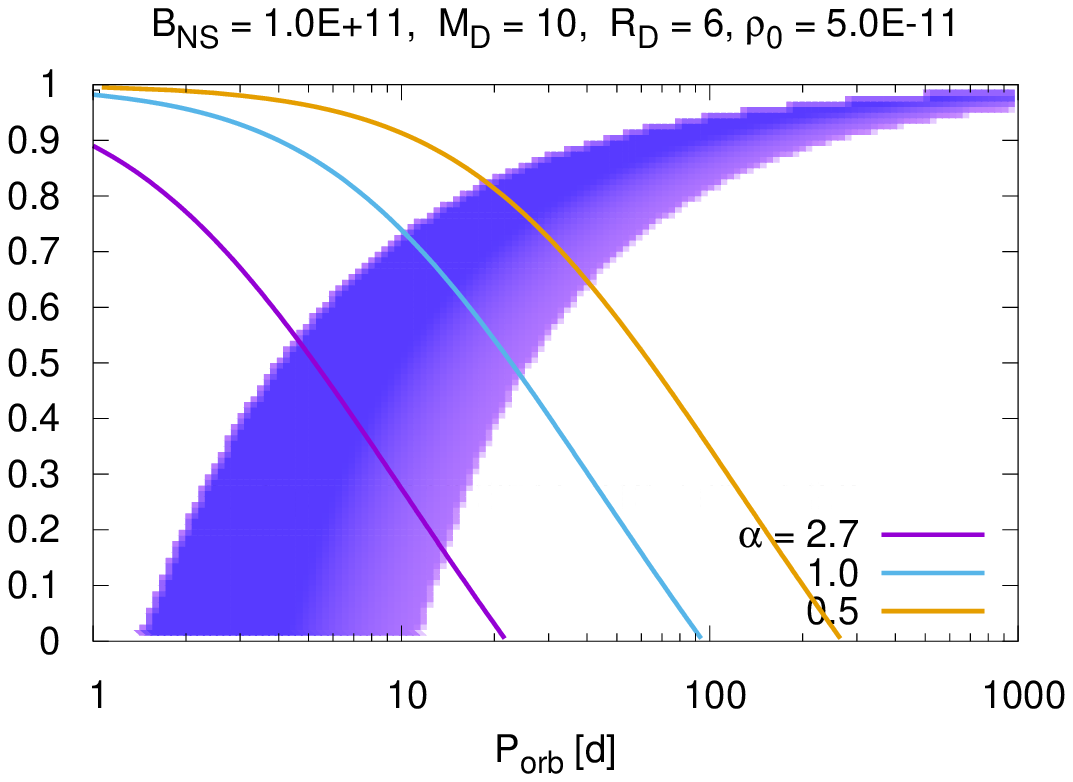}
\includegraphics[width=5.5cm]{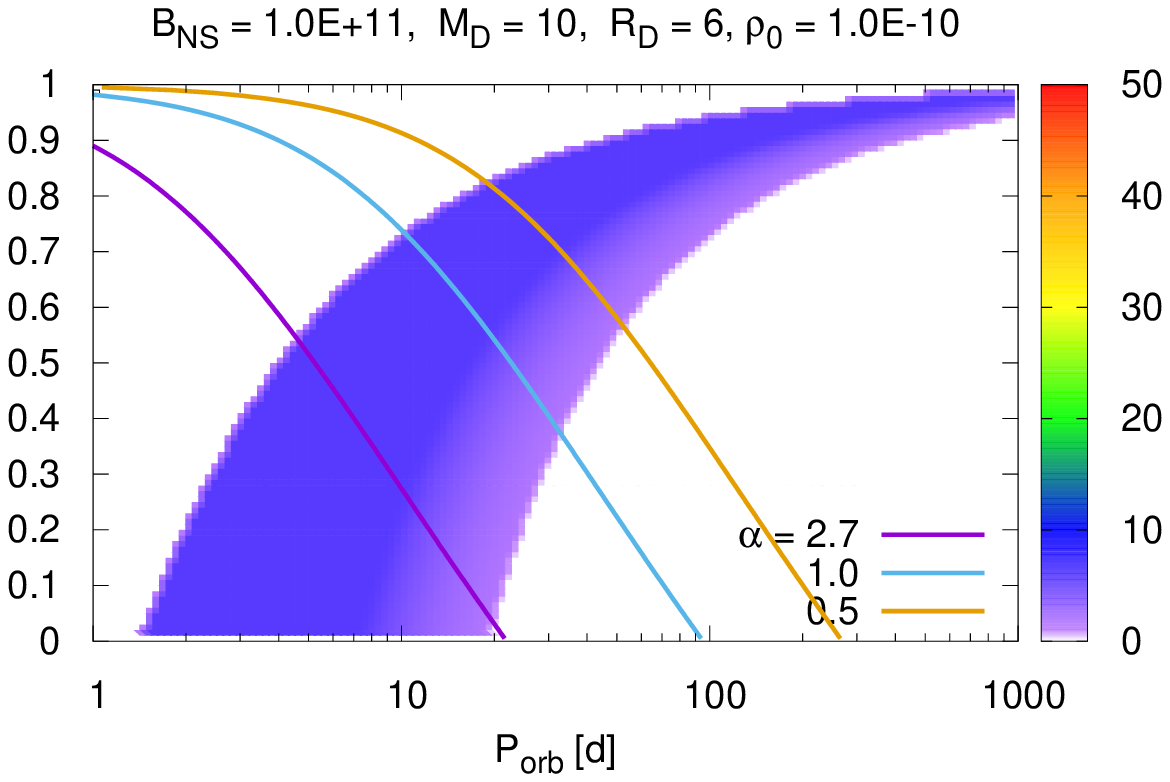} 

\caption{
Region in $P_{\rm{orb}} \text{-} e$ plane where a binary system with a Be donor can be a ULX. 
The colour contours show the maximum possible X-ray luminosity normarized by Eddington luminosity, and the white areas correspond to the areas where the Eddington limit is not reached.
All Be donors in the binary are fixed at $10 M_{\odot}$ and $6 R_{\odot}$, and an inclusion angle is also fixed as $\delta = \pi/8$. 
From left to right in the nine figures, the base density of the Be disc is $\rho_{0} = 2\times 10^{-11}$, $5 \times 10^{-11}$, and $1 \times 10^{-10}\ \rm{g \, cm}^{-3}$. 
The magnetic field of the NS is $B_{\rm{NS}} = 1 \times 10^{13}$, $10^{12}$, and $10^{11}\ \rm{G}$, starting from the top line. 
Given that the determination of $\alpha_{2}$ is rather worse, we have drawn three lines in the diagram to represent the conditions (\ref{eq:warpcondition2}) and which correspond to $\alpha_{2} = 2.7$, 1, and 0.5, from the top to bottom, respectively. 
The region to the lower left of these lines is where the warped Be disc forms.
}
\label{fig:main}
\end{figure*}

Figure~\ref{fig:main} shows the NS luminosity with the orbital period and eccentricity serving as variables and shows the areas in parameter space that satisfy the conditions (\ref{eq:cond1})--(\ref{eq:cond3}).
The colour contour shows the luminosity normalised by the Eddington luminosity.
The mass and radius of the donor are $10 M_{\odot}$ and $6 R_{\odot}$, respectively.
These results reveal a slightly inclined system between the plane of the Be disc and the orbital plane ($\delta = \pi/8$, see Section 4).
Shown are the results given with nine combinations of NS magnetic field and base density of the Be disc. The former varies from $10^{11}$ and $10^{13}\ \rm{G}$ and the latter varies from $10^{-11}$ to $10^{-10}\ \rm{g \, cm}^{3}$.

In figure~\ref{fig:main}, the white area is the region where Be-HMXBs cannot be a ULX because it does not satisfy the conditions mentioned above.
The lower-right region is mainly where the NS orbit does not pass through the high-density region of the Be disc, which means that the Be disc does not transfer enough matter in this region of parameter space.
Conversely, the upper-left region is where the NS orbit is too close to the donor, causing RLOF~\citep{eggleton2006}.
In such a system, if the outer layer of the donor is convective, a common envelope forms and the system will not emit X-rays.
Moreover, the mass ratio becomes large for Be-NS binary systems, in which case a common envelope inevitably forms~\citep{vandenheuvel2017}. 
Although conditions for the formation of the common envelope remain under debate, the binary system causing RLOF may not be a stable system because the orbital parameter changes on a short time scale~\citep{ivanova2013}.

In addition, if the magnetic field is weak, the mass-accretion rate of the NS will be insufficient to transform it into a ULX because the outflow from the disc will dominate.

The base density of the Be disc has a significant effect on the formation of ULXs.
Specifically, a Be disc with a density less than $10^{-11}\ \rm{g \, cm}^3$ cannot produce a sufficiently high-mass-transfer rate to produce a ULX.
However, Be stars with densities greater than $5 \times 10^{-11}\ \rm{g \, cm}^{-3}$ are observed not infrequently~\citep{rivinius2013,touhami2011,draper2014}.

As discussed in the previous section, the luminosity of ULXs increases with NS magnetic field strength.
For a strong magnetic field (${\approx} 10^{13}\ \rm{G}$), the proposed model shows that the source can be up to 50 times brighter than the Eddington luminosity.
However, the brightest systems are limited to narrow regions in parameter space corresponding to short orbital perimeters and small eccentricities.
Even in the parameter region where ULX is possible, the ULX is only moderately bright in many case (i.e. less than 20 times the Eddington limit), which is consistent with the observational suggestion that NS ULXs with Be donors have relatively small luminosity (see Table~\ref{table1}).
Very bright NS ULXs, exceeding $10^{40}\ \rm{erg \, s}^{-1}$, are probably not Be-HMXBs, but rather RLOF systems with supergiant donors, which means that at least two subgroups of NS ULX possibly exist: brighter systems with supergiant donors, and dimmer systems with Be donors.

\subsection{Warp of Be disc}

The misalignment of the orbital plane of the binary system with the Be disc is suggested to be more favourable for the accretion onto the NS in Be-HMXB~\citep{okazaki2013} because, when these planes are aligned, the tidal truncation reduces the size of the Be disc.
As a result, the NS is less likely to gain mass from the Be disc.
If the plane of the Be disc and the orbital plane are misaligned, the outer edge of the Be disc is warped by tidal action.
In addition, the interaction of an NS with this warped region is suggested to increase the mass-transfer rate and lead to a giant outburst in Be-HMXB~\citep{okazaki2013, reig2007, negueruela2001, moritani2011}.

For such mass-transfer from the warped disc to occur, the Be disc radius must be larger than the typical radius at which warping is induced by tidal action.
The radius $r_{\rm{tr}}$ of the truncated disc is taken to be 0.4 to 0.5 times the periastron distance,~\citep{paczynski1977}, so we assume
\begin{equation}
R_{\rm{tr}} = 0.5 a (1-e).
\label{eq:rtr}
\end{equation}

Conversely, the position at which the disc warps due to tidal action is determined by the ratio of the shear viscosity in the azimuthal direction of the Be disc to that in the vertical direction.
Following the work of~\citet{martin2011}, which gives the warping conditions for a Be disc in a binary system, the radius at which the disc warps can be written as
\begin{equation}
\begin{split}
R_{\rm{warp}} & = 4.91 \times 10^{11} \left(1- e^2 \right)^{3/4} \alpha_2^{1/2} h 
\left(\frac{P_{\rm{orb}}}{1 \rm{d}} \right) \\
& \times
\left(\frac{M_{\rm{d}}}{M_{\odot}} \right)^{1/2}
\left(\frac{M_{\rm{NS}}}{M_{\odot}} \right)^{-1/2}
\left(\frac{M_{\rm{d}} + M_{\rm{NS}}}{M_{\odot}} \right)^{1/2}
\left(\frac{R_{\rm{d}}}{R_{\odot}} \right)^{-1/2}\ \rm{[cm]},
\label{eq:rwarp}
\end{split}
\end{equation}
where $h$ is the ratio of the scale height of the disc thickness to the Be stellar radius.
Here, we set $h = 0.04$~\citep{rivinius2013}, and
$\alpha_2$ is the coefficient of the vertical shear viscosity and is given by 
\begin{equation}
\alpha_2 = \frac{1}{2 \alpha_1} \frac{4 \left(1 + 7 \alpha_1^2 \right)}{4 + \alpha_1^2}
\label{eq:alpha}
\end{equation}
where $\alpha_1$ is the dimensionless viscosity parameter \citep{shakura1973,ogilvie1999}.
Numerical results from~\citet{lee1991} suggest that $\alpha_1>0.1$ because an excessively small value of $\alpha_1$ does not provide sufficient mass to the Be disc.
For this reason, $\alpha_1 = 0.3$ is adopted by~\citet{martin2011}, for which eq.~(\ref{eq:alpha}) gives $\alpha_2 = 2.7$.
Conversely,~\citet{cheng2014}, who attempt to explain the existence of giant outbursts in Be-HMXBs based on the warp condition of the Be disc, argue that $\alpha_2 = 0.5 \text{--} 1$ explains the observations.
In this study, we consider $\alpha_2 = 0.5$, 1, and 2.7.

The warping condition of the Be disc in Be-HMXBs is written as
\begin{equation}
R_{\rm{tr}} < R_{\rm{warp}}.
\label{eq:warpcondition}
\end{equation}
Using the Keplerian law in Eq.~(\ref{eq:rtr}) allows the condition (\ref{eq:warpcondition}) to be rewritten as
\begin{equation}
\begin{split}
P_{\rm{orb}} & < 6.91 \times 10^{2} \left[ \frac{ 1-e }{ (1-e^{2})^{3/4} } \right]^{3} \alpha_{2}^{-3/2} \\
 & \times \left(\frac{M_{\rm{d}}}{M_{\odot}} \right)^{-3/2} 
\left(\frac{M_{\rm{d}} + 1.4}{M_{\odot}} \right)^{-1/2}
\left(\frac{R_{\rm{d}}}{R_{\odot}} \right)^{3/2}\ \rm{[d]}.
\label{eq:warpcondition2}
\end{split}
\end{equation}
In this equation, we set the NS mass to $M_{\rm{NS}} = 1.4 M_{\odot}$.

Figure~\ref{fig:main} shows this condition for three different viscosities represented by three different curves that correspond to $\alpha_2 = 0.5$, 1, and 2.7, from top to bottom. 
For each viscosity, the lower part of the curve is where the condition (\ref{eq:warpcondition2}) is satisfied (i.e. where the warped disc forms).
These figures show that, when the vertical viscosity is small ($\alpha_2 = 0.5$), this condition does not significantly affect the conditions for outbursts in the ULX class, and only systems with exceptionally large orbital periods and eccentricities are excluded.
Conversely, for large $\alpha_2$ (${=} 2.7$), as in the case of~\citet{martin2011}, this condition significantly limits the occurrence of giant outbursts, which means that systems with orbital periods of more than a few tens of days and orbital eccentricities of more than 0.3–0.5 are excluded.
The nearby NS ULX systems Swift J0243.6+6124 and SMC X-3, which likely have Be donors, have orbital periods of 28 and 45 d, respectively.
Thus, a large $\alpha_2$ is unlikely to work, and a viscosity of about $\alpha_2 = 1$, as adopted by~\citet{cheng2014}, may be reasonable (see Section 4).

\section{discussion}

In this study, we examine several conditions for Be-HMXBs to occur by mass-accretion beyond the Eddington limit and find the parameter region in which these binary systems become ULXs.
Figure~\ref{fig:main} shows the general results.
Roughly speaking, only Be-HMXBs with small orbital eccentricity and short orbital period can become ULXs.
In the $P_{\rm{orb}} \text{-} e$ plane, a system with a small orbital period suffers RLOF, and such systems are not known to be stable and may form a common envelope~\citep{eggleton2006}.
Conversely, a system with a long orbital period cannot pass through the high-density part of the Be disc if the eccentricity is small, so the mass-transfer rate does not become sufficiently large.
In addition, a system with a long orbital period and a large eccentricity cannot satisfy the warp condition of the Be disc (note that the warp condition depends on the viscosity parameters, which are highly indefinite).
However, in the region where both the orbital period and the eccentricity are large, the probability that a binary system becomes a ULX is relatively low because the allowed ULX region narrows.

When the orbital eccentricity is small, the period of the binary system depends on the base density of the circumstellar disc of the Be donor~\citep{touhami2011, draper2014, rivinius2013}.
In other words, the larger the base density of the Be disc, the more material the NS can acquire and the more likely it is to become a super-Eddington source.
Observationally, the density of the Be disc is about $10^{-10} \text{--} 10^{-12}\ \rm{g \, cm}^{-3}$, and if the disc density is near the upper limit, the mass supply from the Be disc suffices for the NS to become a ULX, and even a binary system with a period of a few tens of days can become a ULX.

The size of the ULX parameter region also depends on the magnetic field of the NS.
If the magnetic field is strong, the condition becomes more stringent for the slim disc to be captured by the magnetic field around the NS (\ref{eq:Mdotsc}), and the ULX parameter region shrinks.
Conversely, if the magnetic radius is large and the accretion disc is captured by the magnetic field of the NS before significant mass loss occurs due to radiation-pressure-induced outflow, both the mass-accretion rate and the X-ray luminosity increase.
Therefore, the X-ray luminosity increases when the NS has strong magnetic field.
As a result, a system with $R_{\rm{sph}}\approx R_{\rm{mag}}$ is likely to be a ULX~\citep{king2017, king2019, king2020}.
However, there is an upper limit to the X-ray luminosity because the ULX parameter region narrows as the magnetic field becomes stronger.
Under the conditions used in our models, the ULX parameter region almost disappears when the neutron star has a magnetar-level magnetic field ($B_{\rm{NS}}\approx 10^{14}\ \rm{G}$).
Therefore, $B_{\rm{NS}} \approx 10^{13}\ \rm{G}$, shown in Fig.~\ref{fig:main}, is almost the upper limit of the magnetic field allowed for a Be-HMXB to become a ULX, and its brightness is limited to about 50 times the Eddington luminosity.
Table~\ref{table1} summarises the donor information and the X-ray luminosities of observed NS ULXs.
The X-ray luminosities of the ULXs with Be-type donors are relatively small, and none of them exceeds 25 times the Eddington limit.
Thus, the proposed model is consistent with observations on this point.

\subsection{Inclination Angle}

A giant outburst is facilitated when the Be disc plane is inclined (i.e. nonparallel) with respect to the orbital plane because the disc is truncated in this case and becomes smaller~\citep{okazaki2013}.
However, the inclination angle between these two planes affects the relative velocity $v_{\rm{rel}}$ of the NS relative to the matter in the Be disc (where $v_{\rm{rel}}$ is the orbital velocity of the NS at the periastron).
Here we assume that the matter in the Be disc rotates around the Be star with Keplerian velocity $v_{\rm{Kep}}$.
Thus, depending on the inclination angle, the relative velocity between the NS and the Be disc matter can vary from $v_{\rm{orb}} - v_{\rm{Kep}}$ to $v_{\rm{orb}} + v_{\rm{Kep}}$. 
We introduce the parameter $\delta$ to represent the inclination angle between the Be disc plane and the orbital plane, and express the relative velocity as 
\begin{equation}
v_{\rm{rel}}^{2} = v_{\rm{orb}}^{2} + v_{\rm{Kep}}^{2} -2 v_{\rm{orb}} v_{\rm{Kep}} \cos \left(\delta \right).
\label{eq:vrel}
\end{equation}
Thus, $\delta$ is the angle between the orbital velocity vector and the Keplerian velocity vector of the disc matter, so $\delta = 0$ corresponds to the coplanar case and $\delta = \pi/2$ corresponds to the disc being perpendicular to the orbital plane.

\begin{figure*}
\includegraphics[width=5.5cm]{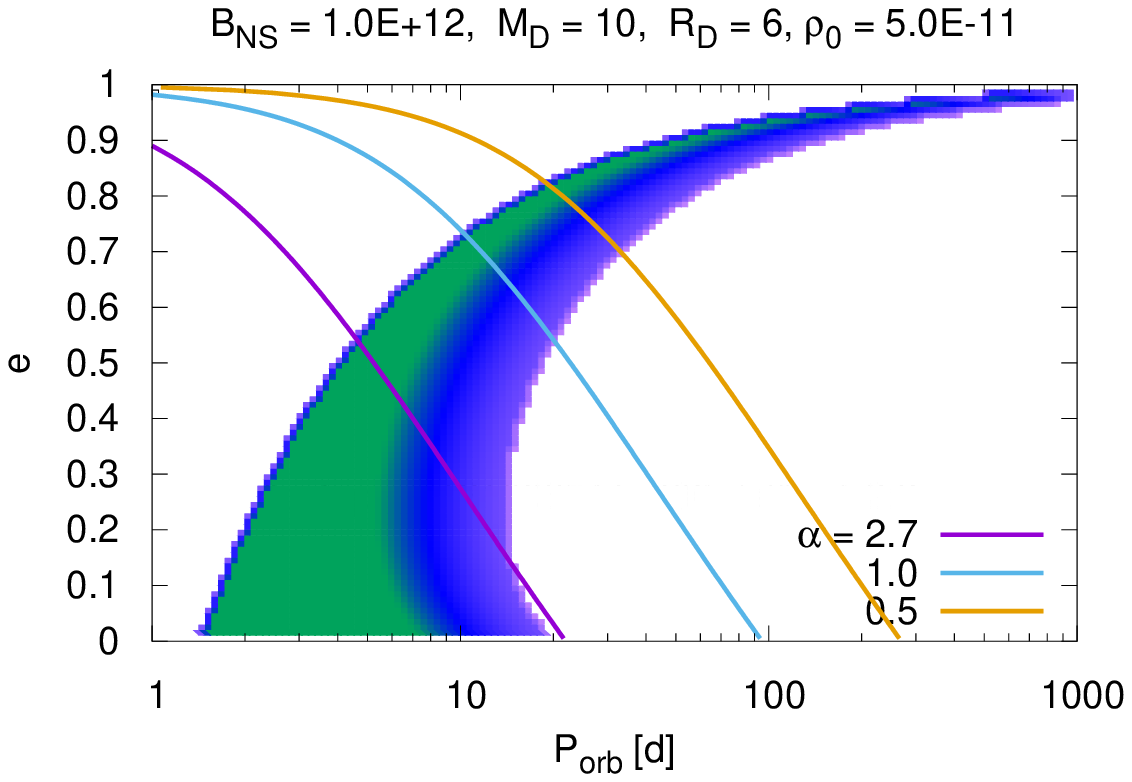}
\includegraphics[width=5.5cm]{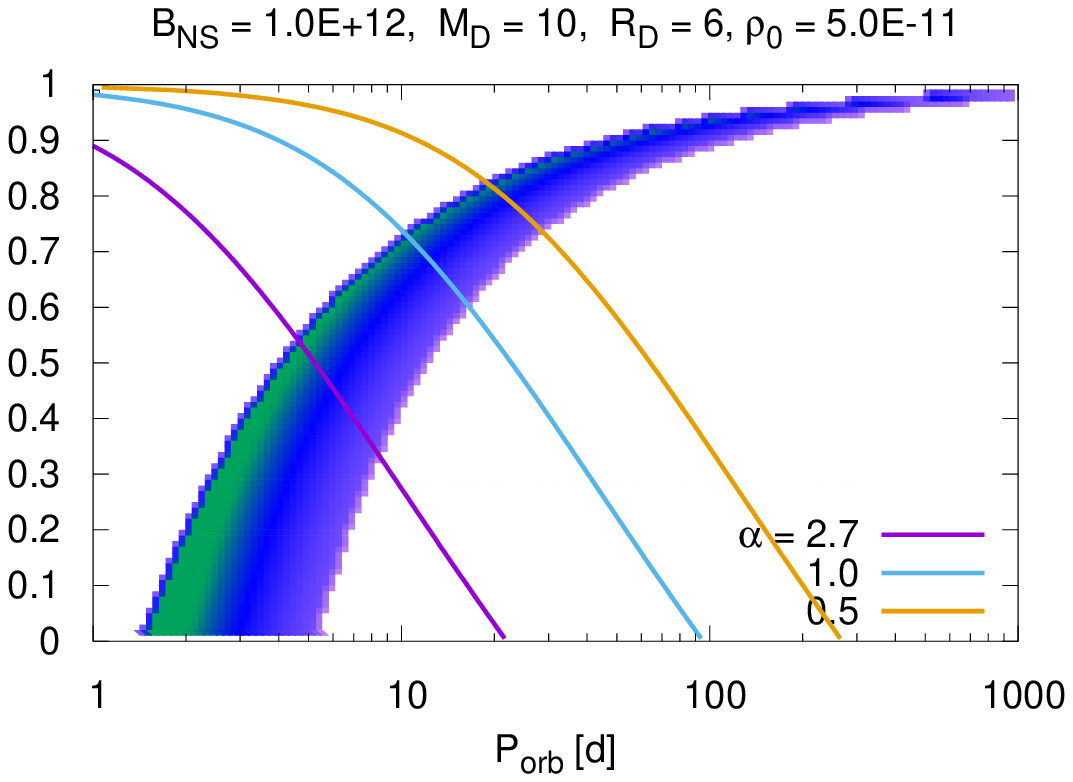}
\includegraphics[width=5.5cm]{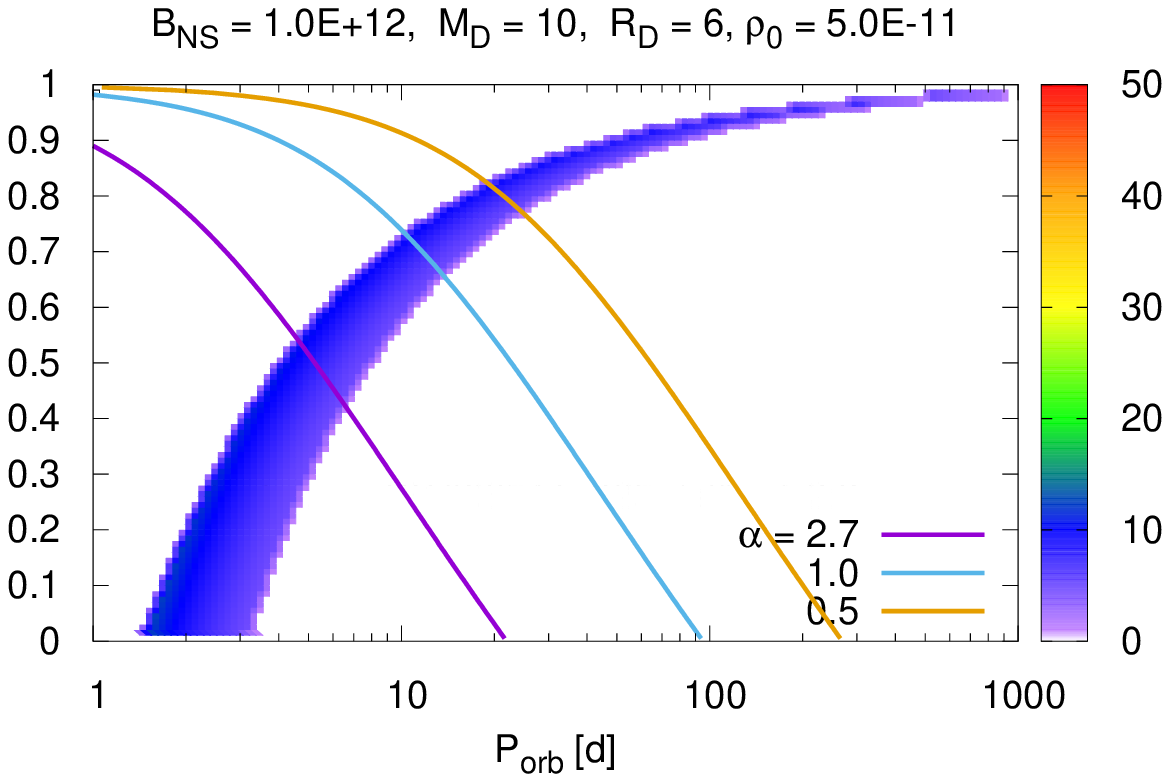}
\caption{
The same diagram as in Fig.~\ref{fig:main} but with different inclusion angles: all Be donors are fixed at $10 M_{\odot}$ and $6 R_{\odot}$, the base density of the Be disc is $\rho_{0} = 5\times 10^{-11}\ \rm{g \, cm}^{-3}$, and the NS magnetic field is $B_{\rm{NS}} = 1\times 10^{12}\ \rm{G}$. 
From left to right, we set $\delta = 0$, $\pi/8$, and $\pi/4$. 
As $\delta$ increases, the relative velocity increases, and the BHL accretion rate decreases, so the ULX parameter region narrows rapidly and disappears from the parameter set.
}
\label{fig:inclination}
\end{figure*}

We now compare the results as a function of inclination angle $\delta$.
Figure~\ref{fig:inclination} shows the results for $B_{\rm{NS}} = 10^{12}\ \rm{G}$ and $\rho_{0} = 5 \times 10^{-11}\ \rm{g, cm}^{-3}$.
Here, as in Fig.~\ref{fig:main}, all Be-type donors are fixed to $10 M_{\odot}$ and $6 R_{\odot}$.
From left to right, figure~\ref{fig:inclination} shows the results for $\delta = 0$, $\pi/8$, and $\pi/4$.
As $\delta$ increases, the relative velocity increases, and the BHL accretion rate decreases, so the ULX parameter region narrows rapidly and disappears from our parameter set.
Conversely, as the relative velocity decreases, the mass-transfer rate due to BHL accretion increases, so the potential ULX parameter region increases.
\citet{okazaki2013} argue that their simulations show that the brightest burst occurs in the coplanar case, despite the largest burst theoretically occurring when the two planes are moderately offset.

\subsection{Conditions Concerning Mass-Transfer Rate}

For a Be-HMXB to become a ULX, this study imposes two conditions concerning the mass-transfer rate: $R_{\rm{mag}} < R_{\rm{sph}}$ and $\dot{M}_{\rm{T}} > \dot{M}_{\rm{sc}}$ \citep{king2017,okazaki2013}.
The former condition is stronger than the latter, so when the former is imposed, the latter can be relaxed, and even if we set $\dot{M}_{\rm{T}}>\dot{M}_{\rm{Edd}}$, this relaxation will have a negligible effect on the results.
For most of NS ULXs where pulsations have been observed, $R_{\rm{mag}} < R_{\rm{sph}}$~\citep{king2017,king2020}.
Therefore, given the observational constraints on spin-up rates and the existence of outflows, we consider it reasonable to impose the condition $R_{\rm{mag}} < R_{\rm{sph}}$ here~\citep{kosec2018, vasilopoulos2021}.

\begin{figure}
\includegraphics[width=\columnwidth]{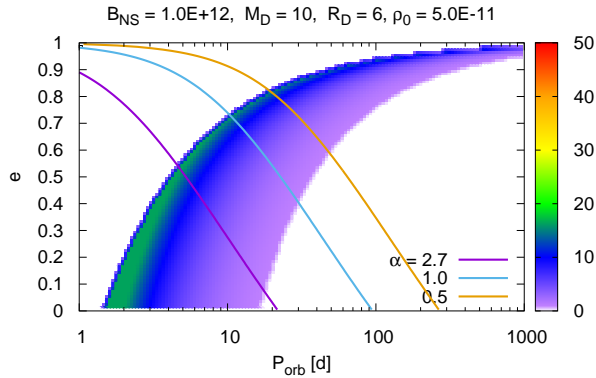}
\caption{
The ULX parameter region when $R_{\rm{mag}} < R_{\rm{sph}}$ and $\dot{M}_{\rm{T}} > \dot{M}_{\rm{sc}}$ are relaxed.
The parameters are the same as those in the centre of Fig. \ref{fig:main}.
When the condition is relaxed, the ULX parameter region expands, but the expanded area corresponds to very dim ULX systems.
}
\label{fig:p}
\end{figure}

The results remains essentially unchanged when the magnetic field is moderate ($10^{12}\ \rm{G}$) and $\dot{M}_{\rm{T}} > \dot{M}_{\rm{sc}}$, even if $R_{\rm{mag}} > R_{\rm{sph}}$.
However, if the magnetic field grows above $10^{12}\ \rm{G}$, 
the ULX parameter region extends significantly.
Figure~\ref{fig:p} shows the results for the standard parameters (i.e. with these two conditions relaxed).
Comparing this with the corresponding results in in the centre of figure~\ref{fig:main}, we see that the ULX parameter region is extended.
However, the majority of the extended parameter region corresponds to a luminosity less than twice the Eddington luminosity of $1.4 M_{\odot}$ objects and is therefore only very faint as a ULX.
In fact, the number of extra-galactic ULXs exceeds $10 L_{\rm{Edd}}$ and, even were the number of fainter objects (${<}10 L_{\rm{Edd}}$) to increase, they would not be observed.
Therefore, for comparison with observations, the results are independent of the conditions relaxed above.

\subsection{Evolution of Binary Systems}

The parameters of a binary system evolve over time.
In particular, due to its relatively high-mass, the radius of a Be donor changes rapidly (${<}1$ Myr).
The ULX parameter region shown in Fig.~\ref{fig:main} depends on the radius of the Be donor.
Figure~\ref{fig:donorradius} shows the evolution of the ULX parameter region as the radius of the Be donor varies from $R_{\rm{d}} = 6$ to $10 R_{\odot}$.
As the Be donor evolves and its radius increases, the potential ULX region clearly broadens and shifts toward the long-$P_{\rm{orb}}$ side.
Toward the end of the main-sequence, when the radius reaches about $10 R_{\odot}$, even a system with an orbital period of about 100 days can become a ULX.
Even a $10 M_{\odot}$ donor loses only about 2\% of its mass during the 10 Myrs of the main-sequence phase, and any change in mass can be neglected during the HMXB phase~\citep{vink2001,hurley2000}. 
Although the maximum accretion rate onto the NS is large, the intermittent nature of accretion in a Be-HMXB means that the cumulative mass-accretion is not large, so changes in NS mass and orbit can be ignored.

The fact that the ULX parameter region expands as the donor evolves means that even a normal Be-type HMXB with a luminosity below the Eddington luminosity can become ULX as the Be donor evolves.
As seen from the Corbet diagram in figure~\ref{fig:CorbetDiagram}, many Be-type HMXBs exist with orbital periods of around 100 d~\citep{klus2014,townsend2011}.
Therefore, a certain fraction of Be-HMXBs may become ULXs during their evolution.
However, numerous unknown features remain, such as whether late-main-sequence Be-type stars retain their discs and how the NS magnetic field evolves.

Additionally, note that HMXBs containing NSs could be progenitors of NS-NS binary systems.
Passing through HMXBs with moderate orbital periods is especially important for the formation of NS-NS binaries that merge within the age of the Universe \citep{belczynski2002,tauris2015}.
In this case, Be-HMXBs with orbital periods ranging from a few days to several hundred days could play an important role as candidate progenitors.
If Be-HMXBs become ULXs that can be observed even outside the galaxy, their population will provide valuable information on the abundance and merger rate of NS-NS binary systems.

\begin{figure*}
\includegraphics[width=5.5cm]{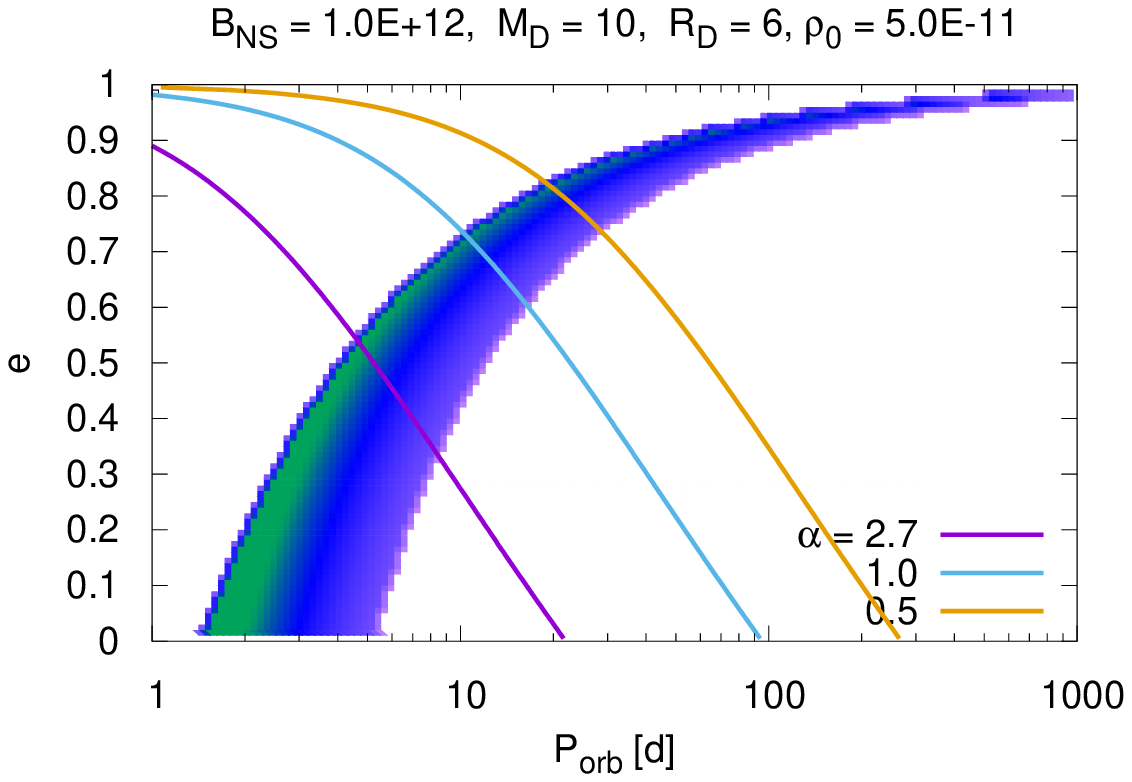}
\includegraphics[width=5.5cm]{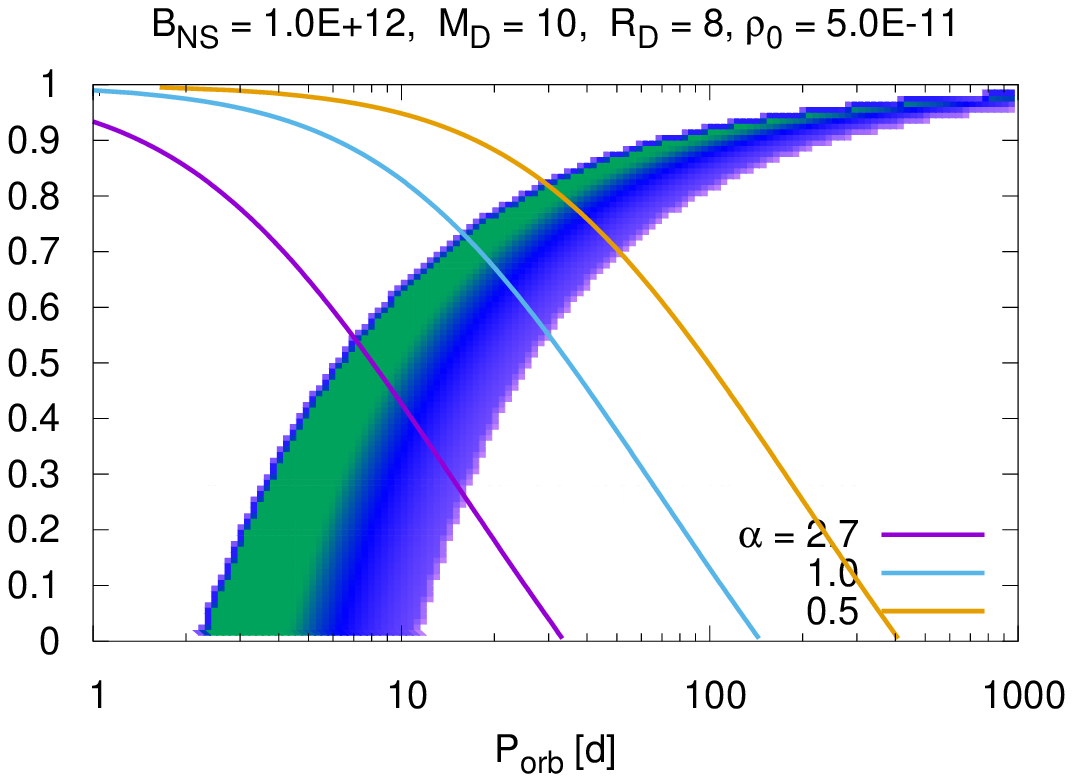}
\includegraphics[width=5.5cm]{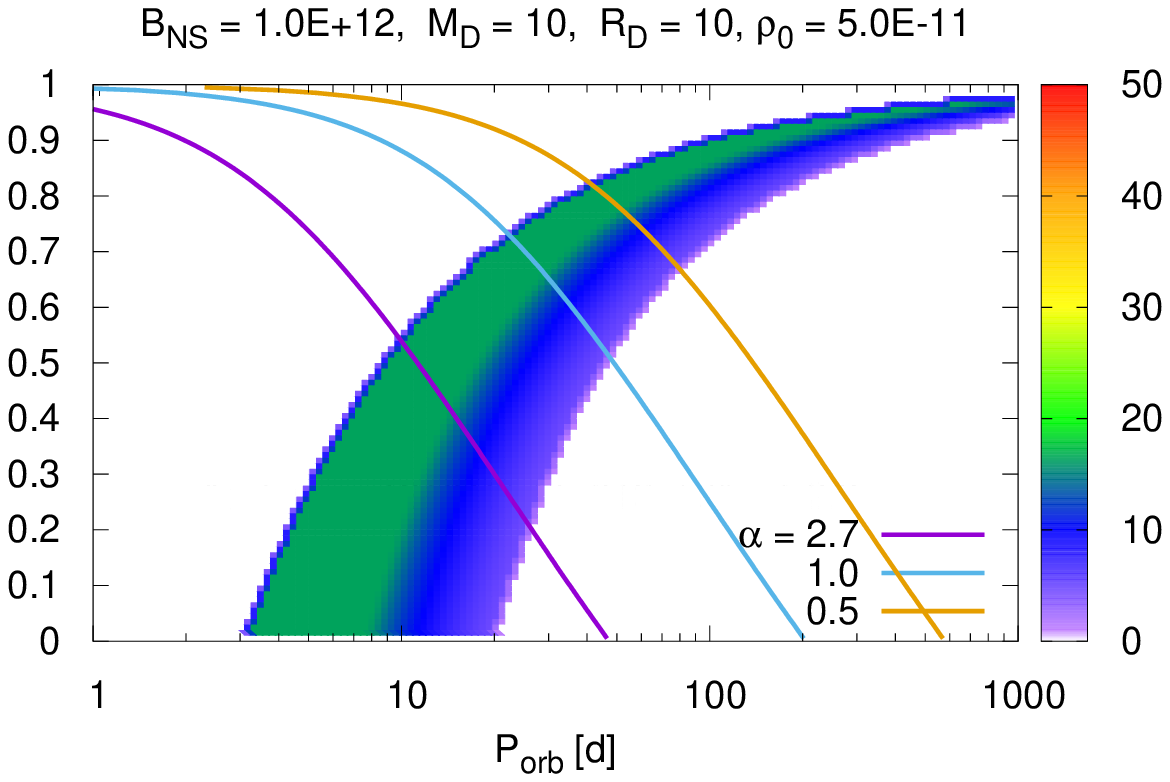}
\caption{
Same diagram as Fig.~\ref{fig:main}, but with different donor radii: the Be-type donor is $10 M_{\odot}$, the base density of the Be disc is $\rho_{0} = 5\times 10^{-11}\ \rm{g \, cm}^{-3}$, the NS magnetic field is $B_{\rm{NS}} = 1\times 10^{12}\ \rm{G }$, and $\delta = \pi /8$. 
The donor radii are $6R_{\odot}$, $8R_{\odot}$, and $10R_{\odot}$, from left to right. 
As the donor evolves and the radius increases, the ULX parameter region widens, and even systems with large orbital periods could become ULXs.
}
\label{fig:rad}
\end{figure*}

\subsection{Comparison with Observed Systems}

\begin{figure}
\includegraphics[width=\columnwidth]{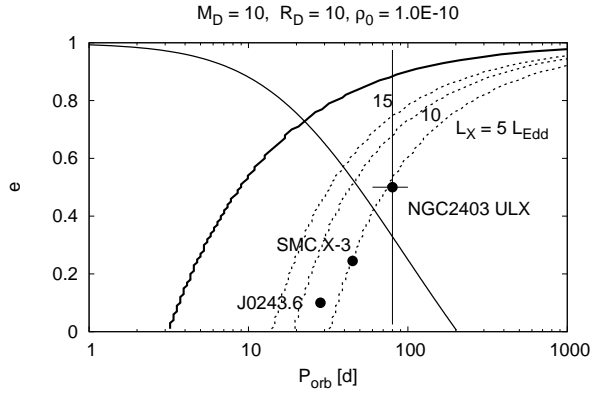}
\caption{
Observed samples are superimposed on the ULX parameter region in a $P_{\rm{orb}} \text{-} e$ diagram. 
Although no data are available on orbital eccentricity for NGC2403 ULX, this figure shows that the eccentricity of this system must be approximately 0.3 or less to enter the ULX parameter region.
}
\label{fig:obs}
\end{figure}

The results obtained are compared with the observed number of NS ULXs. 
Although the magnetic field of the NS and the base density and viscosity of the Be disc may differ in each binary system, we assume the following parameters for the binary systems:
$\rho_{0} = 1 \times 10^{-10}\ \rm{g \, cm}^{-3}$, $M_{\rm{d}} = 10 M_{\odot}$, and $\alpha_{2} = 1.0$.
The parameter region in which a Be-HMXB can exceed the Eddington luminosity and become a ULX is shown by the area enclosed by the curves in figure~\ref{fig:obs}.
The two solid lines indicate the edges of the ULX region in the $P_{\rm{orb}}$ plane (as in Fig.~\ref{fig:main}) and come from the RLOF condition and the warped disc condition.
The three dashed lines correspond to the different X-ray luminosities: $15 L_{\rm{Edd}}$, $10 L_{\rm{Edd}}$, and $5 L_{\rm{Edd}}$, from the left to the right. 
Figure~\ref{fig:obs} also plots the observed NS ULX systems with known orbital periods and which are suggested to have Be donors~\citep{tsygankov2017, doroshenko2018, king2019}.

The luminosity of the NS ULX depends on $B_{\rm{NS}}$ and, for HMXBs, we assume a typical magnetic field strength of $B_{\rm{NS}} = 10^{12}\ \rm{G}$.
Although NS ULXs were initially thought to have a strong magnetic fields~\citep{bachetti2014, eksi2015, mushtukov2015}, recent evidence suggests that this is not necessarily the case~\citep{king2019, middleton2019, mushtukov2019}.
Observational evidence for SMC X-3, shown here as an example, suggests that the magnetic field strength is $B_{\rm{NS}} = (1\text{ to } 5) \times 10^{12}\ \rm{G}$~\citep{tsygankov2017}.

Among the compared systems, SMC X-3 and Swift J0243.6+6124 are in the ULX region.
In the case of these two systems, the BHL process from the warped Be disc transports the material to the vicinity of the NS at a mass-transfer rate above the Eddington rate, and most of this material accretes into the NS magnetosphere.
However, in the case of NGC2404 ULX, when $\alpha_{2} = 1$, the unknown eccentricity must be less than about 0.4 to satisfy the warp condition.
NGC2404 ULX is also likely to have an orbital period close to the lower limit.
Conversely, if the NS magnetic field is small, the ULX parameter region is extended, making it possible for NGX2404 ULX to enter the ULX parameter region even if the orbital period is not so small.

We now consider the X-ray luminosity of these systems when they produce a giant outburst.
Assuming that the magnetic field $B_{\rm{NS}} = 10^{12}\ \rm{G}$, the luminosity of Be-HMXB as a ULX is expected to be at most 10--20 times the Eddington luminosity (cf. Fig.~\ref{fig:main} and Fig.~\ref{fig:obs}).
Even if the magnetic field were a little stronger, by analogy with Fig.~\ref{fig:main}, the X-ray luminosity would be unlikely to exceed 20 times the Eddington luminosity since these sources are located near the right edge of the ULX parameter region. 
However, the luminosity here does not consider how beaming affects the observed luminosity.
The observed luminosity should thus be the intrinsic luminosity divided by the beaming factor $b < 1$~\citep{king2017}.
Some recent studies have argued that beaming is not very strong, so we assume that the effect of beaming is a factor of order unity~\citep{takahashi2017,erkut2020,mushtukov2021}
\footnote{
Recent numerical simulation indicates that beaming could be stronger when $R_{\rm{mag}}$ is much smaller than $R_{\rm{sph}}$ \citep{abarca2021}.
}.
In particular, analyses based on various models suggest that beaming is weak in Swift J2043.6+6124~\citep{erkut2020}.
The X-ray luminosity of observed Be donor systems is 10--20 times the Eddington luminosity (see Table~\ref{table1}), which is consistent with this prediction when we consider weak beaming.

Figure~\ref{fig:main} also shows that the X-ray luminosity of NS ULXs with Be donors does not exceed ${\approx} 50 L_{\rm{Edd}}$ without assuming unreasonable conditions. 
This is also consistent with the fact that the X-ray luminosity of ULXs with Be donors is less than that of ULXs with supergiant donors, as shown in Table~\ref{table1}.

Although the number of observed systems is not large, Table~\ref{table1} shows that most bright NS ULXs have orbital periods of a few days.
Conversely, systems with Be donors have X-ray luminosities less than $20 L_{\rm{Edd}}$ and orbital periods of more than 10 days.
Note that, for another reason, it is unlikely that Be donors exist among the luminous systems with orbital periods of less than 10 days,.
That is, in Be-HMXBs, the NS captures the disc matter as it passes through the Be disc.
Therefore, the donor star must supply the captured mass to the disc.
If the orbital period of the system is too short, it may be impossible to supply and store sufficient mass in the circumstellar disc.
Although an uncertainty exists, the mass supply rate from the Be donor to the circumstellar disc is~\citep{rivinius2013}
$\dot{M}_{\rm{disc}} \approx 10^{18}\ \rm{g \, s}^{-1}$.
Considering that most of the mass is ejected from the outer edge of the disc into interstellar space, it is not surprising that it takes more than 10 days to store sufficient mass in the disc to cause a giant outburst.
A future challenge is to quantify how much mass is transferred from the donor to the disc and from the disc to the NS and interstellar space.

\section{Conclusion}

Recently, several ULXs have been found to have NSs as emitters.
Although the number of observations remains small, some of these ULXs are known to have Be-type donors.
In this paper, we consider the following conditions for a Be-HMXB to become a ULX:
\begin{itemize}
\item[i.] The donor does not suffer from RLOF near the periastron of the binary system.
\item[ii.] The mass capture rate from the Be disc undergoing the BHL process exceeds the Eddington limit.
\item[iii.] Mass-accretion above the Eddington rate proceeds to the NS magnetosphere before massive outflow from the accretion disc.
\end{itemize}
In addition to these criteria, the condition of the Be disc, although highly indefinite, is also considered:
\begin{itemize}
\item[iv.] The Be disc warps inside the truncation radius.
\end{itemize}
We consider that a Be-HMXB satisfying all these conditions should become a NS ULX, and find the requisite parameter region.

The results indicate that a NS should receive mass-accretion above the Eddington rate for a band-like parameter region extending from the lower left to the upper right in the $P_{\rm{orb}}\text{-} e$ plane.
However, the warp conditions of the Be disc do not allow for excessively large eccentricity and orbital period.
The size of the ULX parameter region increases with the base density of the Be disc.
Although the maximum luminosity of the ULX increases as the magnetic field of the NS increases, the ULX parameter region itself decreases as the magnetic field increases, placing an upper limit on the luminosity of NS ULXs with Be-type donors.
Therefore, NS ULXs much brighter than the Eddington luminosity likely form a separate subclass that is subject to RLOF from supergiant donors.

The ULX parameter region also depends on the radius of the Be donor: the larger the radius, the wider the ULX parameter region.
Therefore, as the Be donor evolves and expands, even Be-type HMXBs, whose luminosity is currently below the Eddington luminosity, may evolve into ULXs.
Since Be-HMXBs are quite common, it would be useful to know how many of them become ULXs and for how long during their evolution, so that we can estimate the fraction of ULXs in the observed HMXB population.
If the ULX population can be deduced from the HMXB population, it will also be useful to estimate how many non-pulsating ULXs are of NS origin.
Finally, a general evolutionary model should be constructed for Be-type HMXBs and NS ULXs with Be donors in combination with binary-evolution calculations.

\section*{Data Availability}
 
The data underlying this article will be shared on reasonable request to the corresponding author.

\section*{Acknowledgements}

We would like to thank the reviewer for helpful comments.
This work was supported by JSPS KAKENHI Grant Number 18K03706.




\begin{thebibliography}{99}


\bibitem[\protect\citeauthoryear{{Abarca}, {Parfrey}  \&
  {Klu{\'z}niak}}{{Abarca} et~al.}{2021}]{abarca2021}
{Abarca} D.,  {Parfrey} K.,   {Klu{\'z}niak} W.,  2021, arXiv e-prints,  arXiv:2107.01149

\bibitem[\protect\citeauthoryear{{Abramowicz}, {Czerny}, {Lasota}  \&
  {Szuszkiewicz}}{{Abramowicz} et~al.}{1988}]{abramovicz1988}
{Abramowicz} M.~A.,  {Czerny} B.,  {Lasota} J.~P.,   {Szuszkiewicz} E.,  1988,
 ApJ.332.646

\bibitem[\protect\citeauthoryear{{Aguilera}, {Pons}  \& {Miralles}}{{Aguilera}
  et~al.}{2008}]{auguilera2008}
{Aguilera} D.~N.,  {Pons} J.~A.,   {Miralles} J.~A.,  2008, A\&A.486.255

\bibitem[\protect\citeauthoryear{{Bachetti} et~al.,}{{Bachetti}
  et~al.}{2014}]{bachetti2014}
{Bachetti} M.,  et~al., 2014, Natur.514.202

\bibitem[\protect\citeauthoryear{{Belczynski}, {Kalogera}  \&
  {Bulik}}{{Belczynski} et~al.}{2002}]{belczynski2002}
{Belczynski} K.,  {Kalogera} V.,   {Bulik} T.,  2002, ApJ.572.407

\bibitem[\protect\citeauthoryear{{Bildsten} et~al.,}{{Bildsten}
  et~al.}{1997}]{bildsten1997}
{Bildsten} L.,  et~al., 1997, ApJS.113.367

\bibitem[\protect\citeauthoryear{{Bondi} \& {Hoyle}}{{Bondi} \&
  {Hoyle}}{1944}]{bondi1944}
{Bondi} H.,  {Hoyle} F.,  1944, MNRAS.104.273

\bibitem[\protect\citeauthoryear{{Brightman} et~al.,}{{Brightman}
  et~al.}{2018}]{brightman2018}
{Brightman} M.,  et~al., 2018, NatAs.2.312

\bibitem[\protect\citeauthoryear{{Carpano}, {Haberl}, {Maitra}  \&
  {Vasilopoulos}}{{Carpano} et~al.}{2018}]{carpano2018}
{Carpano} S.,  {Haberl} F.,  {Maitra} C.,   {Vasilopoulos} G.,  2018, MNRAS.476L.45

\bibitem[\protect\citeauthoryear{{Chashkina}, {Lipunova}, {Abolmasov}  \&
  {Poutanen}}{{Chashkina} et~al.}{2019}]{chashkina2019}
{Chashkina} A.,  {Lipunova} G.,  {Abolmasov} P.,   {Poutanen} J.,  2019,
 A\&A.626A.18

\bibitem[\protect\citeauthoryear{{Cheng}, {Shao}  \& {Li}}{{Cheng}
  et~al.}{2014}]{cheng2014}
{Cheng} Z.~Q.,  {Shao} Y.,   {Li} X.~D.,  2014,ApJ.786.128C

\bibitem[\protect\citeauthoryear{{Christodoulou}, {Laycock}, {Yang}  \&
  {Fingerman}}{{Christodoulou} et~al.}{2017}]{christodoulou2017}
{Christodoulou} D.~M.,  {Laycock} S. G.~T.,  {Yang} J.,   {Fingerman} S.,
  2017, RAA.17.59

\bibitem[\protect\citeauthoryear{{Coe} \& {Kirk}}{{Coe} \&
  {Kirk}}{2015}]{coe2015}
{Coe} M.~J.,  {Kirk} J.,  2015, MNRAS.452.969

\bibitem[\protect\citeauthoryear{{Corbet}}{{Corbet}}{1984}]{corbet1984}
{Corbet} R.~H.~D.,  1984, A\&A.141.91

\bibitem[\protect\citeauthoryear{{Corbet}}{{Corbet}}{1986}]{corbet1986}
{Corbet} R.~H.~D.,  1986, MNRAS.220.1047

\bibitem[\protect\citeauthoryear{{Doroshenko}, {Tsygankov}  \&
  {Santangelo}}{{Doroshenko} et~al.}{2018}]{doroshenko2018}
{Doroshenko} V.,  {Tsygankov} S.,   {Santangelo} A.,  2018, A\&A.613A.19

\bibitem[\protect\citeauthoryear{{Draper}, {Wisniewski}, {Bjorkman}, {Meade},
  {Haubois}, {Mota}, {Carciofi}  \& {Bjorkman}}{{Draper}
  et~al.}{2014}]{draper2014}
{Draper} Z.~H.,  {Wisniewski} J.~P.,  {Bjorkman} K.~S.,  {Meade} M.~R.,
  {Haubois} X.,  {Mota} B.~C.,  {Carciofi} A.~C.,   {Bjorkman} J.~E.,  2014,
  ApJ.786.120

\bibitem[\protect\citeauthoryear{{Edgar}}{{Edgar}}{2004}]{edgar2004}
{Edgar} R.,  2004, NewAR.48.843

\bibitem[\protect\citeauthoryear{{Eggleton}}{{Eggleton}}{1983}]{eggleton1983}
{Eggleton} P.~P.,  1983,ApJ.268.368

\bibitem[\protect\citeauthoryear{{Eggleton}}{{Eggleton}}{2006}]{eggleton2006}
{Eggleton} P.,  2006, {Evolutionary Processes in Binary and Multiple Stars}

\bibitem[\protect\citeauthoryear{{Eksi}, {Andac}, {Cikintoglu}, {Gencali},
  {Gungor}  \& {Oztekin}}{{Eksi} et~al.}{2015}]{eksi2015}
{Eksi} K.~Y.,  {Andac} I.~C.,  {Cikintoglu} S.,  {Gencali} A.~A.,  {Gungor} C.,
    {Oztekin} F.,  2015, MNRAS.448L.40

\bibitem[\protect\citeauthoryear{{Erkut}, {T{\"u}rko{\u{g}}lu}, {Ek{\c{s}}i}
  \& {Alpar}}{{Erkut} et~al.}{2020}]{erkut2020}
{Erkut} M.~H.,  {T{\"u}rko{\u{g}}lu} M.~M.,  {Ek{\c{s}}i} K.~Y.,   {Alpar}
  M.~A.,  2020,ApJ.899.97

\bibitem[\protect\citeauthoryear{{Fragos}, {Linden}, {Kalogera}  \&
  {Sklias}}{{Fragos} et~al.}{2015}]{fragos2015}
{Fragos} T.,  {Linden} T.,  {Kalogera} V.,   {Sklias} P.,  2015,ApJ.802L.5

\bibitem[\protect\citeauthoryear{{Frank}, {King}  \& {Raine}}{{Frank}
  et~al.}{2002}]{frank2002}
{Frank} J.,  {King} A.,   {Raine} D.~J.,  2002, {Accretion Power in
  Astrophysics: Third Edition}

\bibitem[\protect\citeauthoryear{{F{\"u}rst} et~al.,}{{F{\"u}rst}
  et~al.}{2016}]{furst2016}
{F{\"u}rst} F.,  et~al., 2016,ApJ.831L.14

\bibitem[\protect\citeauthoryear{{Heida} et~al.,}{{Heida}
  et~al.}{2019}]{heida2019}
{Heida} M.,  et~al., 2019,ApJ.883L.34

\bibitem[\protect\citeauthoryear{{Hoyle} \& {Lyttleton}}{{Hoyle} \&
  {Lyttleton}}{1939}]{hoyle1939}
{Hoyle} F.,  {Lyttleton} R.~A.,  1939, PCPS.35.405

\bibitem[\protect\citeauthoryear{{Hurley}, {Pols}  \& {Tout}}{{Hurley}
  et~al.}{2000}]{hurley2000}
{Hurley} J.~R.,  {Pols} O.~R.,   {Tout} C.~A.,  2000, MNRAS.315.543

\bibitem[\protect\citeauthoryear{{Inoue}, {Ohsuga}  \& {Kawashima}}{{Inoue}
  et~al.}{2020}]{inoue2020}
{Inoue} A.,  {Ohsuga} K.,   {Kawashima} T.,  2020, PASJ.72.34

\bibitem[\protect\citeauthoryear{{Israel} et~al.,}{{Israel}
  et~al.}{2017a}]{israel2017b}
{Israel} G.~L.,  et~al., 2017a, Sci.355.817

\bibitem[\protect\citeauthoryear{{Israel} et~al.,}{{Israel}
  et~al.}{2017b}]{israel2017a}
{Israel} G.~L.,  et~al., 2017b,MNRAS.466L.48

\bibitem[\protect\citeauthoryear{{Ivanova} et~al.,}{{Ivanova}
  et~al.}{2013}]{ivanova2013}
{Ivanova} N.,  et~al., 2013,A\&ARv.21.59

\bibitem[\protect\citeauthoryear{{Kaaret}, {Feng}  \& {Roberts}}{{Kaaret}
  et~al.}{2017}]{kaaret2017}
{Kaaret} P.,  {Feng} H.,   {Roberts} T.~P.,  2017,ARA\&A.55.303

\bibitem[\protect\citeauthoryear{{Karino} \& {Miller}}{{Karino} \&
  {Miller}}{2016}]{karino2016}
{Karino} S.,  {Miller} J.~C.,  2016, MNRAS.462.3476

\bibitem[\protect\citeauthoryear{{Karino}, {Nakamura}  \& {Taani}}{{Karino}
  et~al.}{2019}]{karino2019}
{Karino} S.,  {Nakamura} K.,   {Taani} A.,  2019,PASJ.71.58

\bibitem[\protect\citeauthoryear{{King} \& {Lasota}}{{King} \&
  {Lasota}}{2019}]{king2019}
{King} A.,  {Lasota} J.-P.,  2019, MNRAS.485.3588

\bibitem[\protect\citeauthoryear{{King} \& {Lasota}}{{King} \&
  {Lasota}}{2020}]{king2020}
{King} A.,  {Lasota} J.-P.,  2020, MNRAS.494.3611

\bibitem[\protect\citeauthoryear{{King}, {Lasota}  \& {Klu{\'z}niak}}{{King}
  et~al.}{2017}]{king2017}
{King} A.,  {Lasota} J.-P.,   {Klu{\'z}niak} W.,  2017,MNRAS.468L.59

\bibitem[\protect\citeauthoryear{{Klus}, {Ho}, {Coe}, {Corbet}  \&
  {Townsend}}{{Klus} et~al.}{2014}]{klus2014}
{Klus} H.,  {Ho} W.~C.~G.,  {Coe} M.~J.,  {Corbet} R.~H.~D.,   {Townsend}
  L.~J.,  2014, MNRAS.437.3863

\bibitem[\protect\citeauthoryear{{Kosec}, {Pinto}, {Walton}, {Fabian},
  {Bachetti}, {Brightman}, {F{\"u}rst}  \& {Grefenstette}}{{Kosec}
  et~al.}{2018}]{kosec2018}
{Kosec} P.,  {Pinto} C.,  {Walton} D.~J.,  {Fabian} A.~C.,  {Bachetti} M.,
  {Brightman} M.,  {F{\"u}rst} F.,   {Grefenstette} B.~W.,  2018, MNRAS.479.3978

\bibitem[\protect\citeauthoryear{{Kuranov}, {Postnov}  \&
  {Yungelson}}{{Kuranov} et~al.}{2020}]{kuranov2020}
{Kuranov} A.~K.,  {Postnov} K.~A.,   {Yungelson} L.~R.,  2020, arXiv e-prints,
   arXiv:2010.03488

\bibitem[\protect\citeauthoryear{{Lee}, {Osaki}  \& {Saio}}{{Lee}
  et~al.}{1991}]{lee1991}
{Lee} U.,  {Osaki} Y.,   {Saio} H.,  1991, MNRAS.250.432

\bibitem[\protect\citeauthoryear{{Lipunov}}{{Lipunov}}{1982}]{lipunov1982}
{Lipunov} V.~M.,  1982, SvA.26.54

\bibitem[\protect\citeauthoryear{{Lipunova}}{{Lipunova}}{1999}]{lipunova1999}
{Lipunova} G.~V.,  1999, AstL.25.508

\bibitem[\protect\citeauthoryear{{Martin}, {Pringle}, {Tout}  \&
  {Lubow}}{{Martin} et~al.}{2011}]{martin2011}
{Martin} R.~G.,  {Pringle} J.~E.,  {Tout} C.~A.,   {Lubow} S.~H.,  2011,
  MNRAS.416.2827

\bibitem[\protect\citeauthoryear{{Mart{\'\i}nez-N{\'u}{\~n}ez}
  et~al.,}{{Mart{\'\i}nez-N{\'u}{\~n}ez} et~al.}{2017}]{martinez2017}
{Mart{\'\i}nez-N{\'u}{\~n}ez} S.,  et~al., 2017, SSRv.212.59

\bibitem[\protect\citeauthoryear{{Middleton}, {Brightman}, {Pintore},
  {Bachetti}, {Fabian}, {F{\"u}rst}  \& {Walton}}{{Middleton}
  et~al.}{2019}]{middleton2019}
{Middleton} M.~J.,  {Brightman} M.,  {Pintore} F.,  {Bachetti} M.,  {Fabian}
  A.~C.,  {F{\"u}rst} F.,   {Walton} D.~J.,  2019, MNRAS.486.2

\bibitem[\protect\citeauthoryear{{Misra}, {Fragos}, {Tauris}, {Zapartas}  \&
  {Aguilera-Dena}}{{Misra} et~al.}{2020}]{misra2020}
{Misra} D.,  {Fragos} T.,  {Tauris} T.~M.,  {Zapartas} E.,   {Aguilera-Dena}
  D.~R.,  2020, A\&A.642A.174

\bibitem[\protect\citeauthoryear{{Mondal}, {Belczy{\'n}ski}, {Wiktorowicz},
  {Lasota}  \& {King}}{{Mondal} et~al.}{2020}]{mondal2020}
{Mondal} S.,  {Belczy{\'n}ski} K.,  {Wiktorowicz} G.,  {Lasota} J.-P.,   {King}
  A.~R.,  2020, MNRAS.491.2747

\bibitem[\protect\citeauthoryear{{Moritani}, {Nogami}, {Okazaki}, {Imada},
  {Kambe}, {Honda}, {Hashimoto}  \& {Ichikawa}}{{Moritani}
  et~al.}{2011}]{moritani2011}
{Moritani} Y.,  {Nogami} D.,  {Okazaki} A.~T.,  {Imada} A.,  {Kambe} E.,
  {Honda} S.,  {Hashimoto} O.,   {Ichikawa} K.,  2011, PASJ.63L.25

\bibitem[\protect\citeauthoryear{{Mushtukov}, {Suleimanov}, {Tsygankov}  \&
  {Poutanen}}{{Mushtukov} et~al.}{2015}]{mushtukov2015}
{Mushtukov} A.~A.,  {Suleimanov} V.~F.,  {Tsygankov} S.~S.,   {Poutanen} J.,
  2015, MNRAS.454.2539

\bibitem[\protect\citeauthoryear{{Mushtukov}, {Ingram}, {Middleton}, {Nagirner}
   \& {van der Klis}}{{Mushtukov} et~al.}{2019}]{mushtukov2019}
{Mushtukov} A.~A.,  {Ingram} A.,  {Middleton} M.,  {Nagirner} D.~I.,   {van der
  Klis} M.,  2019, MNRAS.484.687

\bibitem[\protect\citeauthoryear{{Mushtukov}, {Portegies Zwart}, {Tsygankov},
  {Nagirner}  \& {Poutanen}}{{Mushtukov} et~al.}{2021}]{mushtukov2021}
{Mushtukov} A.~A.,  {Portegies Zwart} S.,  {Tsygankov} S.~S.,  {Nagirner}
  D.~I.,   {Poutanen} J.,  2021,MNRAS.501.2424

\bibitem[\protect\citeauthoryear{{Negueruela} \& {Okazaki}}{{Negueruela} \&
  {Okazaki}}{2001}]{negueruela2001}
{Negueruela} I.,  {Okazaki} A.~T.,  2001, A\&A.369.108

\bibitem[\protect\citeauthoryear{{Ogilvie}}{{Ogilvie}}{1999}]{ogilvie1999}
{Ogilvie} G.~I.,  1999, MNRAS.304.557

\bibitem[\protect\citeauthoryear{{Okazaki}, {Hayasaki}  \&
  {Moritani}}{{Okazaki} et~al.}{2013}]{okazaki2013}
{Okazaki} A.~T.,  {Hayasaki} K.,   {Moritani} Y.,  2013, PASJ.65.41

\bibitem[\protect\citeauthoryear{{Paczynski}}{{Paczynski}}{1977}]{paczynski1977}
{Paczynski} B.,  1977,ApJ.216.822

\bibitem[\protect\citeauthoryear{{Poutanen}, {Lipunova}, {Fabrika}, {Butkevich}
   \& {Abolmasov}}{{Poutanen} et~al.}{2007}]{poutanen2007}
{Poutanen} J.,  {Lipunova} G.,  {Fabrika} S.,  {Butkevich} A.~G.,   {Abolmasov}
  P.,  2007, MNRAS.377.1187

\bibitem[\protect\citeauthoryear{{Pringle} \& {Rees}}{{Pringle} \&
  {Rees}}{1972}]{pringle1972}
{Pringle} J.~E.,  {Rees} M.~J.,  1972, A\&A.21.1

\bibitem[\protect\citeauthoryear{{Reig}, {Larionov}, {Negueruela}, {Arkharov}
  \& {Kudryavtseva}}{{Reig} et~al.}{2007}]{reig2007}
{Reig} P.,  {Larionov} V.,  {Negueruela} I.,  {Arkharov} A.~A.,
  {Kudryavtseva} N.~A.,  2007, A\&A.462.1081

\bibitem[\protect\citeauthoryear{{Rivinius}, {Carciofi}  \&
  {Martayan}}{{Rivinius} et~al.}{2013}]{rivinius2013}
{Rivinius} T.,  {Carciofi} A.~C.,   {Martayan} C.,  2013, A\&ARv.21.69

\bibitem[\protect\citeauthoryear{{Rodr{\'\i}guez Castillo}
  et~al.,}{{Rodr{\'\i}guez Castillo} et~al.}{2020}]{rodriguez2020}
{Rodr{\'\i}guez Castillo} G.~A.,  et~al., 2020,ApJ.895.60

\bibitem[\protect\citeauthoryear{{Sathyaprakash} et~al.,}{{Sathyaprakash}
  et~al.}{2019}]{Sathaprakash2019}
{Sathyaprakash} R.,  et~al., 2019, MNRAS.488L.35

\bibitem[\protect\citeauthoryear{{Shakura} \& {Sunyaev}}{{Shakura} \&
  {Sunyaev}}{1973}]{shakura1973}
{Shakura} N.~I.,  {Sunyaev} R.~A.,  1973, A\&A.24.337

\bibitem[\protect\citeauthoryear{{Shao} \& {Li}}{{Shao} \&
  {Li}}{2015}]{shao2015}
{Shao} Y.,  {Li} X.-D.,  2015,ApJ.802.131

\bibitem[\protect\citeauthoryear{{Takahashi} \& {Ohsuga}}{{Takahashi} \&
  {Ohsuga}}{2017}]{takahashi2017}
{Takahashi} H.~R.,  {Ohsuga} K.,  2017, ApJ.845L.9

\bibitem[\protect\citeauthoryear{{Tauris}, {Langer}  \&
  {Podsiadlowski}}{{Tauris} et~al.}{2015}]{tauris2015}
{Tauris} T.~M.,  {Langer} N.,   {Podsiadlowski} P.,  2015, MNRAS.451.2123

\bibitem[\protect\citeauthoryear{{Tong}}{{Tong}}{2015}]{tong2015}
{Tong} H.,  2015, RAA.15.517

\bibitem[\protect\citeauthoryear{{Touhami}, {Gies}  \& {Schaefer}}{{Touhami}
  et~al.}{2011}]{touhami2011}
{Touhami} Y.,  {Gies} D.~R.,   {Schaefer} G.~H.,  2011, ApJ.729.17

\bibitem[\protect\citeauthoryear{{Townsend}, {Owocki}  \& {Howarth}}{{Townsend}
  et~al.}{2004}]{townsend2004}
{Townsend} R.~H.~D.,  {Owocki} S.~P.,   {Howarth} I.~D.,  2004,MNRAS.350.189

\bibitem[\protect\citeauthoryear{{Townsend}, {Coe}, {Corbet}  \&
  {Hill}}{{Townsend} et~al.}{2011}]{townsend2011}
{Townsend} L.~J.,  {Coe} M.~J.,  {Corbet} R.~H.~D.,   {Hill} A.~B.,  2011,
 MNRAS.416.1556

\bibitem[\protect\citeauthoryear{{Tsygankov}, {Doroshenko}, {Lutovinov},
  {Mushtukov}  \& {Poutanen}}{{Tsygankov} et~al.}{2017}]{tsygankov2017}
{Tsygankov} S.~S.,  {Doroshenko} V.,  {Lutovinov} A.~A.,  {Mushtukov} A.~A.,
  {Poutanen} J.,  2017, A\&A.605A.39

\bibitem[\protect\citeauthoryear{{Vasilopoulos}, {Koliopanos}, {Haberl},
  {Treiber}, {Brightman}, {Earnshaw}  \& {G{\'u}rpide}}{{Vasilopoulos}
  et~al.}{2021}]{vasilopoulos2021}
{Vasilopoulos} G.,  {Koliopanos} F.,  {Haberl} F.,  {Treiber} H.,  {Brightman}
  M.,  {Earnshaw} H.~P.,   {G{\'u}rpide} A.,  2021, arXiv e-prints, arXiv:2102.07996

\bibitem[\protect\citeauthoryear{{Vink}, {de Koter}  \& {Lamers}}{{Vink}
  et~al.}{2001}]{vink2001}
{Vink} J.~S.,  {de Koter} A.,   {Lamers} H.~J.~G.~L.~M.,  2001, A\&A.369.574

\bibitem[\protect\citeauthoryear{{Walton} et~al.,}{{Walton}
  et~al.}{2018}]{walton2018}
{Walton} D.~J.,  et~al., 2018,ApJ.857L.3

\bibitem[\protect\citeauthoryear{{Wang}}{{Wang}}{1981}]{wang1981}
{Wang} Y.~M.,  1981, A\&A.102.36

\bibitem[\protect\citeauthoryear{{Weng}, {Ge}, {Zhao}, {Wang}, {Zhang}, {Bian}
  \& {Yuan}}{{Weng} et~al.}{2017}]{weng2017}
{Weng} S.-S.,  {Ge} M.-Y.,  {Zhao} H.-H.,  {Wang} W.,  {Zhang} S.-N.,  {Bian}
  W.-H.,   {Yuan} Q.-R.,  2017, ApJ.843.69

\bibitem[\protect\citeauthoryear{{Wilson-Hodge} et~al.,}{{Wilson-Hodge}
  et~al.}{2018}]{wilson2018}
{Wilson-Hodge} C.~A.,  et~al., 2018, ApJ.863.9

\bibitem[\protect\citeauthoryear{{Zhang} \& {Kojima}}{{Zhang} \&
  {Kojima}}{2006}]{zhang2006}
{Zhang} C.~M.,  {Kojima} Y.,  2006, MNRAS.366.137

\bibitem[\protect\citeauthoryear{{van den Heuvel}, {Portegies Zwart}  \& {de
  Mink}}{{van den Heuvel} et~al.}{2017}]{vandenheuvel2017}
{van den Heuvel} E.~P.~J.,  {Portegies Zwart} S.~F.,   {de Mink} S.~E.,  2017,
  MNRAS.471.4256

\end{thebibliography}






i


\bsp	
\label{lastpage}
\end{document}